\documentclass[fleqn,usenatbib]{mnras}
\usepackage{newtxtext,newtxmath}
\usepackage[T1]{fontenc}

\DeclareRobustCommand{\VAN}[3]{#2}
\let\VANthebibliography\thebibliography
\def\thebibliography{\DeclareRobustCommand{\VAN}[3]{##3}\VANthebibliography}

\usepackage{graphicx}	% Including figure files
\usepackage{amsmath}	% Advanced maths commands
\usepackage{threeparttable}
\usepackage{xcolor}
\usepackage{multicol}

\newcommand{\oeight}{[O\,{\sc iii}]$\,88\,\micron$}
\newcommand{\ofive}{[O\,{\sc iii}]$\,52\,\micron$}
\newcommand{\ctwo}{[C\,{\sc ii}]$\,158\,\micron$}

\newcommand{\oiii}{[O\,{\sc iii}]$\lambda$}
\newcommand{\hei}{He\,{\sc i}~$\lambda$}

\newcommand{\oii}{[O\,{\sc ii}]$\lambda$}
\newcommand{\neiii}{[Ne\,{\sc iii}]$\lambda$}

\newcommand{\htwo}{H\,{\sc ii}}

\title[JWST NIRCam and NIRSpec study of a galaxy at $z=7.212$]{RIOJA. Young Starburst and Ionized Gas Outflows in a $z = 7.212$ Galaxy Uncovered by JWST NIRCam and NIRSpec Observations}
\author[Y. W. Ren et al.]{
Yi W. Ren,$^{1}$\thanks{E-mail: renyi@toki.waseda.jp}
Akio K. Inoue,$^{1,2}$
Javier \'Alvarez-M\'arquez,$^{3}$
Takuya Hashimoto,$^{4,5}$
Luis Colina,$^3$
\newauthor~Yuma Sugahara,$^{1,2,6}$
Luca Costantin,$^3$
Ken Mawatari,$^{2,4,5}$
Yoshinobu Fudamoto,$^{2,6,7}$
Santiago Arribas,$^3$
\newauthor~Alejandro Crespo G\'omez,$^3$
Daniel Ceverino,$^8$
Yurina Nakazato,$^9$
Masato Hagimoto,$^{10}$
Mitsutaka Usui,$^4$
\newauthor~Rui Marques-Chaves,$^{11}$
Hiroshi Matsuo,$^{6,12}$
Takeshi Hashigaya,$^{13}$
Wataru Osone,$^4$
Carmen Blanco-Prieto,$^3$
\newauthor~Yoichi Tamura,$^{10}$
Naoki Yoshida,$^{9,14,15}$
Tom J. L. C. Bakx$^{16}$
and Miguel Pereira-Santaella$^{17}$
\\
$^{1}$Department of Physics, Graduate School of Advanced Science and Engineering, Faculty of Science and Engineering, Waseda University, 3-4-1 Okubo, Shinjuku, \\~~~Tokyo 169-8555, Japan \\
$^{2}$Waseda Research Institute for Science and Engineering, Faculty of Science and Engineering, Waseda University, 3-4-1 Okubo, Shinjuku, Tokyo 169-8555, Japan\\
$^{3}$Centro de Astrobiolog\'{\i}a (CAB), CSIC-INTA, Ctra. de Ajalvir km 4, Torrej\'on de Ardoz, E-28850, Madrid, Spain \\
$^4$Division of Physics, Faculty of Pure and Applied Sciences, University of Tsukuba, Tsukuba, Ibaraki 305-8571, Japan\\
$^5$Tomonaga Center for the History of the Universe (TCHoU), Faculty of Pure and Applied Sciences, University of Tsukuba, Tsukuba, Ibaraki 305-8571, Japan\\
$^6$National Astronomical Observatory of Japan, 2-21-1 Osawa, Mitaka, Tokyo 181-8588, Japan\\
$^7$Center for Frontier Science, Chiba University, 1-33 Yayoi-cho, Inage-ku, Chiba 263-8522, Japan\\
$^8$Universidad Autonoma de Madrid, Ciudad Universitaria de Cantoblanco, E-28049 Madrid, Spain\\
$^9$Department of Physics, The University of Tokyo, 7-3-1 Hongo, Bunkyo, Tokyo 113-0033, Japan\\
$^{10}$Department of Physics, Graduate School of Science, Nagoya University, Nagoya, Aichi 464-8602, Japan\\
$^{11}$Geneva Observatory, Department of Astronomy, University of Geneva, Chemin Pegasi 51, CH-1290 Versoix, Switzerland\\
$^{12}$The Graduate University for Advanced Studies (SOKENDAI), 2-21-1 Osawa, Mitaka, Tokyo 181-8588, Japan\\
$^{13}$Department of Astronomy, Kyoto University, Sakyo-ku, Kyoto 606-8502, Japan\\
$^{14}$Kavli Institute for the Physics and Mathematics of the Universe (WPI), UT Institute for Advanced Study, The University of Tokyo, Kashiwa, Chiba 277-8583, Japan\\
$^{15}$Research Center for the Early Universe, School of Science, The University of Tokyo, 7-3-1 Hongo, Bunkyo, Tokyo 113-0033, Japan\\
$^{16}$Department of Space, Earth and Environment, Chalmers University of Technology, Onsala Space Observatory, SE-439 92 Onsala, Sweden\\
$^{17}$Instituto de F\'isica Fundamental (IFF), CSIC, Serrano 123, E-28006, Madrid, Spain
}

\date{Accepted XXX. Received YYY; in original form ZZZ}

\pubyear{\the\year{}}

\begin{document}
\label{firstpage}
\pagerange{\pageref{firstpage}--\pageref{lastpage}}
\maketitle

\begin{abstract}
We present analysis of JWST NIRCam and NIRSpec observations of the galaxy SXDF-NB1006-2 at $z = 7.212$, as part of the Reionization and the ISM/Stellar Origins with JWST and ALMA (RIOJA) project. 
We derive the physical properties by conducting spectral energy distribution (SED) fitting, revealing that our target is a young (age $\sim2$ Myr) starburst galaxy with intense radiation field.
We detect multiple nebular emission lines from NIRSpec IFS data. We identify a robust broad component of \oiii5008 emission, indicating the presence of ionized gas outflows. 
The derived gas depletion time of a few hundred Myr implies that our target could be one of the progenitors of massive quiescent galaxies at $z\sim4-5$ identified by recent JWST observations.
The spatial distribution of optical and far-infrared (FIR) [O\,{\sc iii}] emission lines differs in morphology, likely resulting from different critical densities and inhomogeneous density distributions within the galaxy.
Potential old stellar populations may be necessary to account for the derived metallicity of $\sim0.2\,\rm{Z}_\odot$, and their presence can be confirmed by future MIRI observations.
Including our target, star-forming galaxies at $z>6$ detected by ALMA are generally very young but more massive and brighter in UV than galaxies identified by only JWST. The ALMA-detected galaxies may also have a steeper mass-metallicity relation. These findings suggest that the ALMA-detected galaxies may have experienced more efficient mass assembly processes in their evolutionary pathways.

\end{abstract}


\section{Introduction}
Understanding properties of stellar components and the interstellar medium (ISM) in galaxies at $z>6$ is crucial for studying cosmic reionization, galaxy formation and evolution. The improvement of observational instruments has pushed the frontiers of extragalactic studies to more distant Universe and more detailed structures within galaxies. Ground-based telescopes such as Subaru Telescope and Keck Telescope are equipped with near-infrared (NIR) instruments to observe rest-frame UV and optical light from distant galaxies. However, their sensitivities are affected by atmospheric absorption and their resolutions are limited by atmospheric seeing if not set up with Adaptive Optics. As the most powerful space telescope to date, James Webb Space Telescope (JWST; \citealt{gardner23}) has advanced NIR and mid-infrared (MIR) capabilities with better sensitivity, resolution and wavelength coverage compared to previous infrared space telescopes such as Spitzer and NIR devices onboard Hubble Space Telescope (HST). It has enabled the investigation of the first galaxies in the Universe \citep[e.g.,][]{robertson22NA, Donnan23,Harikane23,curtis-lake23,bingjie23_uncover_z13,hainline24_jades_z>8, carniani24_2,heintzDLA, zavalaMIRI, robertson_jades_z14}. Recent JWST observations also have revealed star-forming galaxies in the early Universe with extremely young stellar ages \citep[e.g.,][]{Heintz_z8, tang23, chen23, Endsley23faintuv, looser23,robertson22NA, sugahara25}, ionized gas outflows \citep[e.g.,][]{tang23, carniani24}, increasing galaxy merger rates with redshift up to $z\sim6$ \citep{duan24}, increasing electron densities in ISM with redshift \citep[][]{isobe23, abdurro24, zavalaMIRI} and mildly evolved mass-metallicity relation at $z>3$ \citep[e.g.,][]{nakajima23, curti24, venturi24}.

Additionally, radio interferometer arrays such as the Atacama Large Millimeter/submillimeter Array (ALMA) and the Northern Extended Millimeter Array (NOEMA) provide observational windows in radio wavelengths, enabling the studies of far-infrared (FIR) emission lines (e.g., [O\,{\sc iii}], [C\,{\sc ii}], [N\,{\sc ii}]) and dust continuum from galaxies in the reionization epoch (e.g., \citealt{inoue16,carniani17,smit18, hashimoto18,hashimoto19,Tamura19, Bakx20,ha20,ca20,sugahara21, fudamoto21,bouwens22, zavalaFIRO3, fudamoto_noema, carniani_FIRO3, schouws25, ishii24}). Early galaxies identified by radio telescopes are unique targets for JWST follow-up studies. \citet{inoue16, ha20} found ALMA-detected [O\,{\sc iii}] emitters at $z>6$ exhibit higher [O\,{\sc iii}]/[C\,{\sc ii}] luminosity ratios, suggesting that $z>6$ galaxies may have different ISM conditions and properties compared to local sources. Furthermore, \citet{ha20} reported that $z>6$ galaxies may have a steeper slope of the [C\,{\sc ii}]-star formation rate (SFR) correlation compared to local dwarf galaxies, while \citet{ca20} and \citet{schaerer20} found no or little evolution of the [C\,{\sc ii}]-SFR relation across the cosmic time. With high sensitivities, JWST spectroscopic observations can provide complementary information about line and continuum emission, gas distribution and kinematics from rest-UV to NIR ranges of faint high-$z$ galaxies. Thus, ALMA and JWST joint analysis is critical for understanding the physical properties of ISM and how they evolve across the cosmic time.

The research target of this work, SXDF-NB1006-2 at $z=7.212$, was the most distant galaxy known in 2012. Subaru/Suprime-Cam NB1006 narrow band imaging and Keck/DEIMOS spectroscopy first identified its Ly$\alpha$ emission line \citep{shibuya12}. Subaru/Suprime-Cam $z'$ band, UKIRT/WFCAM $J, H, K$ bands, and Spitzer $3.6\,\micron$ and $4.5\,\micron$ bands also observed the rest-UV and optical emission in this galaxy. However, except for $J$ band, only non-detections were obtained \citep{inoue16}. They performed SED fitting based on these photometries and identified SXDF as a young starburst galaxy with extremely short star formation timescale of $1-2$ Myr, high SFR of $\log (\rm{SFR} / \textit{M}_\odot\,\rm{yr}^{-1}) = 2.54^{+0.17}_{-0.71}$, stellar mass of $\log (M_\ast/\rm{M}_\odot) = 8.54^{+0.79}_{-0.22}$, little dust attenuation of $E(B - V) = 0-0.04$ mag, possible low metallicity of $0.05 - 1\,\rm{Z}_\odot$ and ionizing photon escape fraction of $0-71$ per cent. Consequently, it may be one of the typical star-forming galaxies (SFGs) that contribute to the cosmic reionization. Nevertheless, these values were estimated using many non-detections and suffer from large uncertainties. JWST photometries with high sensitivities are capable of deriving these physical properties more precisely. 

Moreover, SXDF-NB1006-2 was the first galaxy from which the \oeight~emission was detected with ALMA \citep{inoue16}. \citet{ca20} found a tentative detection of \ctwo~emission from this galaxy by analyzing archival ALMA data. 
\citet{yi23} analyzed follow-up ALMA observations targeting \oeight~with high angular resolution and identified a clumpy structure of the \oeight~emission. The dust continuum in ALMA band 6 and 8 was reported as a non-detection in the above literature. Being equipped with high spatial resolution, JWST imaging and integral field unit (IFU) spectroscopy are able to unveil the morphology of emission from rest-UV to NIR wavelengths with unprecedented details. It is time to reveal the nature of this interesting target in the epoch of reionization.

This paper is organized as follows. The observations of our target, data reduction and NIRCam photometries are described in Section 2. The analysis of SED fitting is shown in Section 3. The analysis of emission lines obtained from NIRSpec data is presented in Section 4. The discussion is presented in Section 5. The conclusion is given in Section 6. In this paper, we assume $\textit{H}_{0}$ = 70 km s$^{-1}$ Mpc$^{-1}$, $\Omega_{M}$ = 0.3 and $\Omega_{\Lambda}$ = 0.7. In this case, $1\arcsec$ corresponds to 5.134 kpc at $z=7.212$. Vacuum wavelengths are shown for emission lines. This paper adopts the initial mass function (IMF) from \citet{Kroupa2001}.

\section{data}
\subsection{Observation and Reduction}
The NIRCam and NIRSpec observations were carried out as part of the Reionization and the ISM/Stellar Origins with JWST and ALMA (RIOJA) project (JWST GO1 PID 1840; PIs: J. \'Alvarez-M\'arquez and T. Hashimoto; \citealt{hashimoto23, sugahara25}). 

The NIRCam observations were conducted on August 16, 2022, covering $1-5\,\micron$ in observed frame. The effective exposure time for NIRCam filters are 816 seconds (F115W and F277W), 2104 seconds (F150W and F444W) and 730 seconds (F200W and F356W). NIRCam imaging data were calibrated using a custom strategy based on the JWST calibration pipeline version 1.12.3 \citep{nircam_pipeline} and CRDS context 1145. The calibration included snowballs and wisps removal adopting the strategy described in \citet{bagley23} and further background homogenization applying the same strategy as in \citet{Perez-Gonzalez23}. All images were resampled to a common pixel scale of 0\farcs03/pixel.

The NIRSpec integral field spectroscopy (IFS) observations were conducted on July 24, 2023. The data were taken with a four-point MEDIUM cycling dither pattern and the grating/filter pair of G395M/F290LP, covering 2.87--5.27\,\micron~in the observed frame with the medium spectral resolution (\(R\sim1000\)).
Neither leakcal nor off-scene background observations were conducted as our project targets emission lines of galaxies.
The effective exposure time was 1284 seconds.
The data reduction was based on the JWST public pipeline version 1.11.4 \citep{nirspec_pipeline} under CRDS context jwst\_1118.pmap.
We applied additional reduction processes, including
(1) the \(1/f\)-noise removal,
(2) the rejection of hot pixels and cosmic-ray effects by sigma clipping,
(3) the mask of pixels recording the flux leakage from the failed open MSA shutter, and
(4) the global background subtraction by moving average as a function of wavelengths.
The final output cube was constructed with drizzle weighting to have the spatial scales of 0\farcs05/pixel. The spectral resolutions at wavelengths of \oiii$\lambda$4960, 5008, H$\beta$, H$\gamma$, H$\epsilon$, \neiii3870 and \oii$\lambda$3727, 3730 lines are $R = 1042, 1032, 1011$,901, 824, 804 and 773, respectively, taken from the dispersion and resolution fits for G395M disperser in JWST User Documentation \footnote{\url{https://jwst-docs.stsci.edu}}.

We utilized 3 ALMA datasets to make continuum map of band 6 (rest-$160\,\micron$; 2012.1.00374.S, 2013.A.00021.S, 2019.1.01634.L), 2 datasets to make continuum map of band 8 (rest-$90\,\micron$; 2013.1.01010.S, 2015.A.00018.S) and 1 dataset to make continuum map of band 9 (rest-$50\,\micron$; 2021.1.01323.S). The data reduction and imaging were conducted by a standard way described in \citet{yi23} and Ren et al. in preparation.

\subsection{Correction for Astrometry}
\label{astromety}
\begin{figure*}
\includegraphics[width=\textwidth]{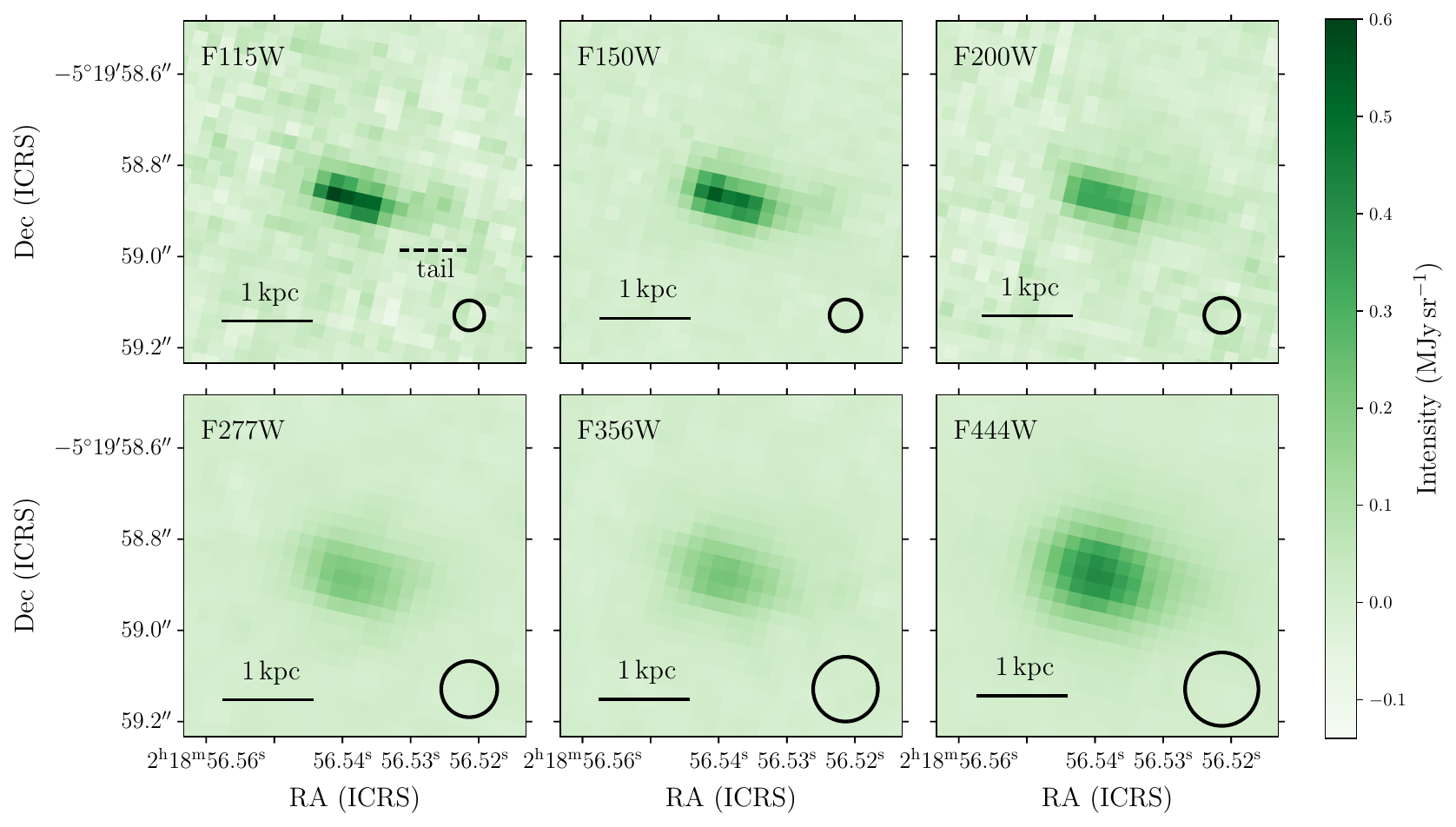}
\caption{NIRCam images of SXDF-NB1006-2. The black circle in the bottom right corner of each panel indicates the PSF FWHM taken from \citet{Finkelstein}. Above the dashed line in F115W represents the tail structure.} \label{image}  
\end{figure*}

We corrected the astrometry of NIRCam images using  stars from Gaia DR3 catalog in the observed field of view (FoV) \citep{gaiamission, gaiadr3, gaiacatalogue}. There are two Gaia stars in total. We performed 2D Gaussian fitting using Photutils \citep{bradley22} to measure their centroids in NIRCam images, and compared the measured centroids with their coordinates recorded in Gaia Archive. The proper motions of Gaia stars were corrected. The offsets between the measured centroids and the archived coordinates differ across filters. Concretely speaking, the offsets of Right Ascension (RA) and Declination (Dec) in short wavelength (SW) images are $0\farcs03-0\farcs04$~and $0\farcs06-0\farcs08$, respectively. While the offsets of RA and Dec in long wavelength (LW) images are $0\farcs002-0\farcs01$ and $0\farcs03-0\farcs04$, respectively. Thus, we independently corrected the astrometry for each band's image by applying the measured average offset of RA and Dec. However, in F150W's image, a bad pixel obscured the peak position of a Gaia star, preventing accurate centroid measurement. Therefore, we aligned the F150W astrometry by matching the centroids of other two compact sources between F150W and astrometry-corrected F200W images. The NIRCam images of the target galaxy after correcting for the astrometry are shown in Fig. \ref{image}. 

We corrected the astrometry of NIRSpec IFU data by aligning the centroid of \oiii$\lambda4960, 5008$ emission with the centroid of astrometry-corrected F444W image, because the emission in F444W is predominantly contributed by [O\,{\sc iii}].

\subsection{Aperture Photometry of NIRCam data}
\label{aperphot}
From Figure \ref{image}, the UV continuum detected in F115W, F150W, and F200W is elongated toward the southwest direction, and it shows a tail-like structure. The emission in LW images looks more extended, which may result from lower resolutions in LW images. The emission in F444W is also more luminous and extended than the optical continuum in F277W and F356W, due to the strong emission from optical [O\,{\sc iii}] doublet lines.

We performed aperture photometry to measure the flux density of our target in each filter using Photutils \citep{bradley22}. Because of the extended structure of the emission in each filter, we performed curve of growth analysis to measure the total flux density. We determined the largest radius of the circular aperture to measure the total flux density at the point where its growth curve approximately levels off. This approach allows us to measure the total flux experimentally, thus no further aperture correction was applied. The determined aperture sizes are shown in Table \ref{photometric measuements}. The aperture size varies across different filters due to the differing spatial scales of emission observed in each filter. To determine photometric uncertainties, we performed random aperture photometry, considering possible spatial correlations of noise between pixels. We masked our target and bright noisy pixels in this process. For each filter, the aperture size of random apertures is the same as that to measure the total flux density. Our measurements are listed in Table \ref{photometric measuements}. 
\begin{table}
\centering
\caption{Photometric measurements in NIRCam images.}
\label{photometric measuements}
\begin{threeparttable}
\begin{tabular}{cccc}
\hline
Band & Radius (\arcsec)\tnote{$^\ast$} & $f_{\nu}~(1\times10^{-7}~\rm{Jy})$ & S/N \\
\hline
F115W & 0.30 & $2.47\pm0.27$ & 9.3  \\
F150W & 0.30 & $2.55\pm0.07$ & 34.2 \\
F200W & 0.30 & $2.00\pm0.21$ & 9.4  \\
F277W & 0.39 & $2.45\pm0.12$ & 19.9 \\
F356W & 0.39 & $2.51\pm0.12$ & 21.3 \\
F444W & 0.81 & $6.25\pm0.15$ & 43.1 \\
\hline
\end{tabular}
\begin{tablenotes}
\item Note:
\item[$^\ast$] Radius of the circular aperture to measure the total flux in every filter. The aperture radius in F444W is much larger because the emission in F444W is more extended than in other filters, due to the H$\beta+$[O\,{\sc iii}] emission. (Fig. \ref{image}).
\end{tablenotes}
\end{threeparttable}
\end{table}

\section{SED fitting}
\label{sed}
\subsection{Data Input}
We performed SED fitting using the Bayesian Analysis of Galaxies for Physical Inference and Parameter EStimation (BAGPIPES) code \citep{bag18}. We input the JWST NIRCam photometric results, the ALMA dust continuum non-detections at bands 6, 8, and 9 measured in \citet{yi23} and Ren et al. in preparation, flux of \oeight~emission \citep{yi23} and flux of optical [O\,{\sc iii}] doublet measured in \S\ref{line}. 
We did not input flux of tentatively detected \ctwo~emission because BAGPIPES only considers nebular emission from \htwo~regions.
We made pseudo narrow band filters to input line fluxes in BAGPIPES. For \oiii$\lambda$4960, 5008, we input their total fluxes, including the broad component of \oiii5008 (\S\ref{double gauss fit}). For \oeight, we input flux measured from $5.9\,\sigma$ detection \citep{yi23}. Since the resolution for photometric output in BAGPIPES is $R = 200$ \citep{bag18}, the wavelength widths of pseudo band transmission curves for optical and FIR [O\,{\sc iii}] are set to be $10^3$ and $10^5\,$\AA, respectively, to ensure that the input lines can be resolved by the spectral resolution. Therefore, the input flux densities are calculated over the widths of pseudo band transmission curves, instead of their real line widths. The input values are $4.2\pm0.2\,\mu\rm{Jy}$ and $0.14\pm0.04\,\rm{mJy}$, for optical and FIR [O\,{\sc iii}], respectively. 

We set a 10 per cent error floor for NIRCam photometries with detections $>10\,\sigma$, to account for possible systematic errors. However, we do not set a 10 per cent error floor for input flux density of optical [O\,{\sc iii}], because this value is one order of magnitude higher than those of NIRCam photometries, and a 10 per cent error will introduce too large uncertainty. To input information of dust continuum non-detections, we set the input fluxes to be 0, and the input flux uncertainties to be their $3\,\sigma$ upper limits.

According to the photometric curve of growth, the half-light scale of UV continuum in F115W, F150W and F200W is $0\farcs18\times0\farcs18$. Previously, we used 1 beam area to calculate the $3\,\sigma$ upper limits of the flux densities of undetected dust continuum \citep{yi23}. If we assume the dust continuum has the same extent as the UV continuum, the angular resolutions of ALMA band 6 ($1\farcs2\times0\farcs77$) and band 8 ($0\farcs17\times0\farcs15$) observations are low enough to detect all the potential flux of the dust continuum \citep{yi23}. However, ALMA band 9 observation has a resolution of $0\farcs15\times0\farcs12$ (Ren et al. in preparation), which is likely to cause flux loss if we measure the flux of dust continuum in band 9 under this resolution. Therefore, we performed \textit{uv}-taper with $0\farcs07$~when making continuum map of ALMA band 9 to scale down its resolution to the half-light scale of UV continuum. The obtained resolution is $0\farcs18\times0\farcs14$, and the measured  $3\,\sigma$ upper limit of flux density is $<0.24\,\rm{mJy}$. This beam area is large enough to cover the total flux of the potential dust continuum, and we adopted the upper limit after \textit{uv}-taper in SED fitting. The information of ALMA measurements input in SED fitting is summarized in Table \ref{alma measurements}.

\begin{table}
    \centering
    \caption{Summary of measurements from ALMA input in SED fitting.}
    \label{alma measurements}
    \begin{threeparttable}
    \begin{tabular}{ccc}
    \hline
         & Beam Size & $F_{\rm{\nu}}$ (mJy) \\
        \hline
        Band 6 & $1\farcs20\times0\farcs77$ & $<0.02~(3\,\sigma)$ \\
        Band 8 & $0\farcs17\times0\farcs15$ & $<0.07~(3\,\sigma)$ \\
        Band 9 & $0\farcs18\times0\farcs14$ & $<0.24~(3\,\sigma)$ \\
        \oeight & $0\farcs17\times0\farcs16$ & $0.14\pm0.04$\tnote{$^\ast$} \\
        \hline
    \end{tabular}
    \begin{tablenotes}
        \item Note:
        \item[$^\ast$]Calculated over the velocity width of the pseudo band transmission curve, instead of the integral width to make moment-0 map in \citet{yi23}.
    \end{tablenotes}
    \end{threeparttable}
\end{table}

\subsection{Parameter Setting}
\label{parameter setting}
In this fitting, we adopted stellar population models from \citet{ssp}, dust emission models from \citet{draine&li}, IGM attenuation model from \citet{inoue14}, dust attenuation law from \citet{calzetti} and Cloudy photoionization code \citep{cloudy} assembled in BAGPIPES. The original fitting range of the ionization parameter in BAGPIPES is from $-4$ to $-2$. We updated its upper limit to 0 with version 22.02 of Cloudy. We set the maximum age of birth clouds to be $0.01\,\rm{Gyr}$ and the multiplicative factor $\eta$ on dust attenuation for stars in birth clouds to be 1. Furthermore, for simplicity, the escape fraction of ionizing photons is assumed to be zero in BAGPIPES. This is different from the previous SED fitting method of the same target in \citet{inoue16}, in which the escape fraction of ionizing photons was a free parameter.

We tried five parametric star formation history (SFH) models (constant, $\tau$ (exponential decline), delayed $\tau$, lognormal and double power-law) and one non-parametric SFH model from \citet{leja19} with a continuity prior in the SED fitting. For non-parametric SFH model, the value of the oldest age bin influences both the time when SF starts and the stellar mass. Therefore, we adopted an age boundary of 20 Myr, which is slightly larger than, but still close to the SF start points derived from lognormal and double power law models. Notably, this start point is different from the SF onset time, where the latter is defined as the time when the SFR exceeds 0.1 per cent of the peak SFR.

\subsection{SED Fitting Results}
\begin{figure*}
    \includegraphics[width = \textwidth]{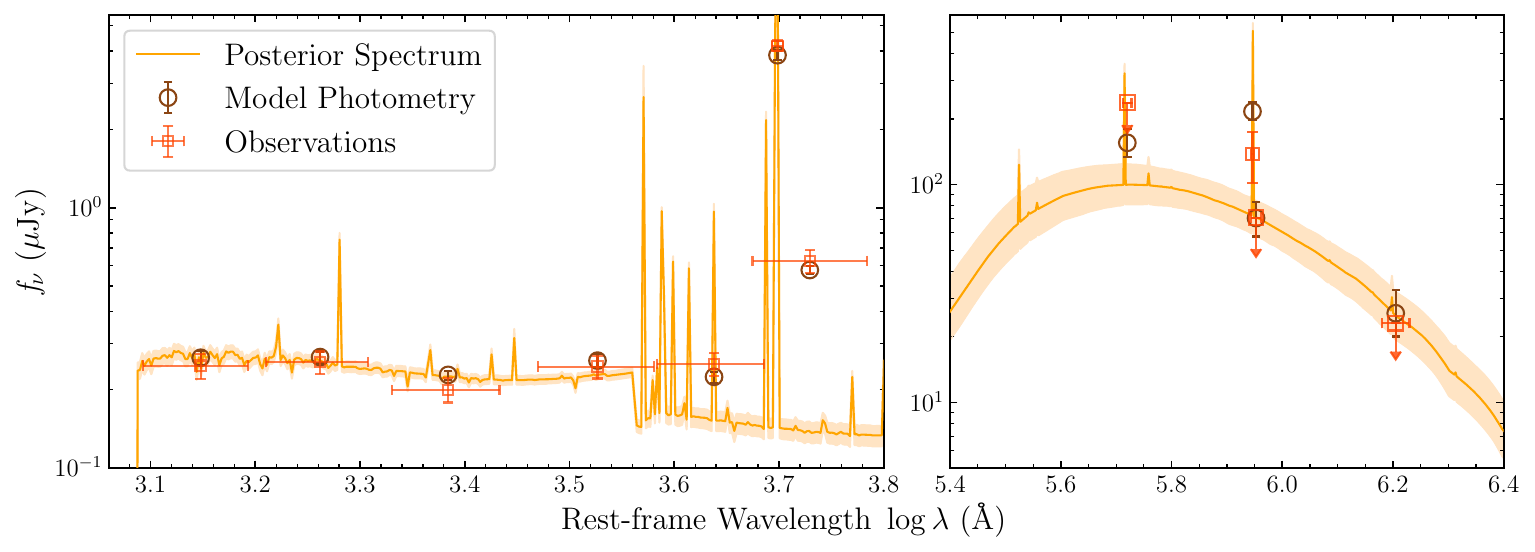}
    \caption{SED fit for SXDF-NB1006-2 with Lognormal SFH model. The horizontal axis shows the rest-frame wavelength. The light orange line represents the 50th percentile of the posterior spectrum output from the SED fitting, and the orange shaded region shows its $1\,\sigma$ deviation (16th-84th percentiles). The dark orange empty squares are the observational data and the brown empty circles are the model photometry outputs from SED fitting. The horizontal error bars of the observational data points represent the wavelength coverage of the filters. Left panel illustrates fitting in rest-UV and optical range with JWST measurements. Right panel shows results in rest-FIR range with ALMA measurements.}
    \label{optical and IR SED}
\end{figure*}

We derived best-fit results of physical quantities and their $1\,\sigma$ uncertainties from the posterior probability distribution of the output samples from SED fitting. We present the best-fit results from six SFH models in Table \ref{SED results} and their best-fit SFHs are supplemented in Fig. \ref{sfh}. Aside from non-parametric model yielding an instantaneous SFR $>2\times$ higher than other models, the best-fit results of SFR, $M_\ast$, $Z$, $A_{\rm{V}}$ and $\log U$ derived from six SFH models are generally consistent with each other. The mass-weighted mean stellar ages across all six SFH models remain similar at $\sim1-2$ Myr as well. However, the ages since onset of SF differ: constant, $\tau$ and delayed $\tau$ models resulted in comparable ages of $\sim2$ Myr, whereas double power law, lognormal and non-parametric models yielded ages 4-5 times larger, as these models incorporate clear rising SF components. There is also a discrepancy between observed and modeled flux in F356W from $\tau$, delayed $\tau$, constant and non-parametric models, while this discrepancy is not present within uncertainties in double power law and lognormal fits. Thus, we adopted the best-fit results from lognormal SFH model as fiducial values, and we show its SED plot in Fig. \ref{optical and IR SED}. The generated SED in FIR range is consistent with the $3\,\sigma$ upper limits of dust continuum non-detections.

\begin{table*}
    \centering
    \caption{Best-fit SED fitting results of phyiscal quantities from different assumptions of SFH models.}
    \label{SED results}
    \begin{threeparttable}
        \renewcommand{\arraystretch}{1.2}
        \begin{tabular}{ccccccccc}
        \hline
         SFH model & SFR\tnote{$^\ast$} & $\log M_\ast$ & $Z$ & $A_{\rm{V}}$ & $t_{\rm{on}}$\tnote{$^\dagger$} & $<t>_{\rm{mass}}\tnote{$^\ddagger$}$ & $\log U$ & $\chi^{2}_\nu$\tnote{$^{\dagger\dagger}$} \\
          & $(M_\odot\,\rm{yr}^{-1})$ & $(\log M_\odot)$ & $(\rm{Z}_\odot)$ & (mag) & (Myr) & (Myr) & & \\
        \hline
        Constant & $206^{+68}_{-58}$ & $8.54^{+0.05}_{-0.04}$ & $0.20^{+0.03}_{-0.01}$ & $0.15\pm0.05$ & $1.7^{+0.6}_{-0.5}$ & $1.1^{+0.2}_{-0.1}$ & $-1.6^{+0.2}_{-0.3}$ & $4.7^{+1.2}_{-0.7}$ (8) \\
        $\tau$ & $193^{+75}_{-51}$ & $8.53^{+0.05}_{-0.04}$ & $0.21\pm0.02$ & $0.14\pm0.06$ & $1.8^{+0.6}_{-0.5}$ & $1.1^{+0.2}_{-0.1}$ & $-1.7^{+0.3}_{-0.2}$ & $7.4^{+2.1}_{-1.2}$ (9) \\
        Delayed $\tau$ & $219^{+74}_{-55}$ & $8.54\pm0.05$ & $0.20^{+0.03}_{-0.01}$ & $0.15\pm0.06$ & $2.2^{+1.0}_{-0.9}$ & $1.1^{+0.2}_{-0.1}$ & $-1.6^{+0.2}_{-0.3}$ & $7.5^{+1.6}_{-1.2}$ (9) \\
        Double power law & $122^{+18}_{-19}$ & $8.59^{+0.05}_{-0.04}$ & $0.20^{+0.02}_{-0.01}$ & $0.23^{+0.06}_{-0.05}$ & $8.2_{-0.4}^{+0.8}$ & $1.9^{+0.1}_{-0.0}$ & $-1.5^{+0.2}_{-0.3}$ & $18.0^{+4.2}_{-3.3}$ (10) \\
        Lognormal\tnote{$^\intercal$} & $165^{+29}_{-21}$ & $8.58^{+0.05}_{-0.04}$ & $0.20^{+0.02}_{-0.01}$ & $0.21\pm0.06$ & $10.4^{+1.6}_{-2.2}$ & $1.8^{+0.1}_{-0.2}$ & $-1.6^{+0.3}_{-0.2}$ & $8.4^{+2.1}_{-1.6}$ (9) \\
        Non-parametric\tnote{$^a$} & $476^{+203}_{-182}$ & $8.58^{+0.06}_{-0.05}$ & $0.20^{+0.02}_{-0.01}$ & $0.18^{+0.06}_{-0.05}$ & $9.9^{+9.92}_{-7.9}$ & $1.3^{+0.3}_{-0.2}$ & $-1.6^{+0.3}_{-0.2}$ & $13.9^{+3.3}_{-2.3}$ (10) \\
        \hline
        \end{tabular}
        \begin{tablenotes}
         \item Note:
         \item The results represent the 50th percentile of posterior output samples, while the uncertainties correspond to the 16th and 84th percentiles, i.e., the $1\,\sigma$ deviations. 
         \item[$^\ast$]Instantaneous SFR at the most recent 1 Myr backwards from the time of observation. 
         \item[$^\dagger$]Age since onset of star formation. The onset of SF is defined as the time when the SFR exceeds 0.1 per cent of the peak SFR.
         \item[$^\ddagger$]Mass-weighted mean age defined by: $<t>_{\rm{mass}} = \int_0^t t \cdot \rm{SFR}(\textit{t})\,d\textit{t}\,/\int_0^t\rm{SFR}(\textit{t})\,d\textit{t}$, where $t$ is the time backwards from the time of observation.
         \item[$^\intercal$]Results from lognormal model are adopted as fiducial values.
         \item[$^{\dagger\dagger}$]Reduced chi-square values. Values in the parentheses indicate the numbers of free parameters. The number of data points for all fittings is 11.
         \item[$^a$]\citet{leja19} SFH model with a continuity prior. The earliest age bin is set at 20 Myr ago.
        \end{tablenotes}
    \end{threeparttable}
\end{table*}

According to the best-fit results, this galaxy has a low dust attenuation, which is constrained by the non-detections of dust continuum. It also has a short star-forming timescale and a high SFR. The spectrum analysis in \S\ref{line} will show there is no evidence of AGN activity in this galaxy. The SFR, even the smaller one measured from H$\beta$ emission ($\log(\rm{SFR}/M_\odot\,\rm{yr}^{-1})\sim1.62$; \S\ref{sfr from hb}), is higher than the SFR of $\log(\rm{SFR}/M_\odot\,\rm{yr}^{-1})=0.8\pm0.4$ for star formation main sequence at $z=7-10$ at the stellar mass of our target ($\sim10^{8.58}\,\rm{M}_\odot$) \citep{Heintz23NA}, confirming that our galaxy is a young starburst. The high ionization parameter of $\log U\sim-1.6$ also implies that the radiation sources in this galaxy are dominated by massive young stars which emit strong ionizing photons. These results are consistent with the blue UV color. We derived the UV slope by fitting the UV continuum inferred from the posterior distribution of SED fitting with a power law $f_\lambda \varpropto \lambda^\beta$ and adopting fitting windows from \citet{calzetti1994}. The obtained $\beta$ is $-2.32\pm 0.04$, consistent with $z\gtrsim6$ galaxies detected by JWST (\citealt{Endsley23faintuv, cullen23,  topping24, cullen24}). However, our galaxy is brighter in rest-UV than these JWST samples ($M_{\rm{UV}}=-22$; see \S\ref{ionization and ism condition}), and the UV slope of our galaxy is bluer than the extrapolation at $M_{\rm{UV}}=-22$ for UV-fainter sources at similar redshifts (median $\beta\sim-2.0$; \citealt{topping24}).

Furthermore, the modeled spectrum from SED fitting exhibits a prominent Balmer jump at rest-frame $3646$\,\AA~(Fig. \ref{optical and IR SED}). Although there is no such feature appealing in the NIRSpec spectrum because it is too close to the lowest detectable wavelength of the adopted NIRSpec filter and the continuum is not detected (\S\ref{line}), the predicted Balmer jump suggests the presence of strong nebular continuum and the radiation sources are dominated by young stellar populations. SED fitting results from recent JWST/NIRCam photometries have also revealed predictions of Balmer jumps and nebular continuum from galaxies at very high redshifts (\citealt{topping24, endsley_jades, sugahara25}). There have also been JWST/NIRSpec observations of LAEs at $z\sim6$ showing Balmer jumps and nebular continuum, and these systems are likely powered by young stellar populations such as Wolf-Rayet stars and metal-poor massive stars (\citealt{Roberts-Borsani24, Cameron24}). We quantified the Balmer jump strength by measuring the flux ratio between F277W and F356W filters, while excluding the contamination from emission lines detected in NIRSpec data (\S\ref{line}). The result is $1.58\pm0.14$, similar to the prediction from a two-photon emission model that successfully explained a UV turnover feature observed at $z\sim6$ \citep{Cameron24}. Such strong two-photon emission is likely powered by very hot stars with effective temperatures of $\sim10^5\,$K \citep{Cameron24}. With the current data, we cannot confirm if there is strong two-photon emission in our target, but future observations will be crucial to examine the presence of strong two-photon emission and very hot stars.

\section{line measurements}
\label{line}
\begin{figure*}
    \includegraphics[width=\textwidth]{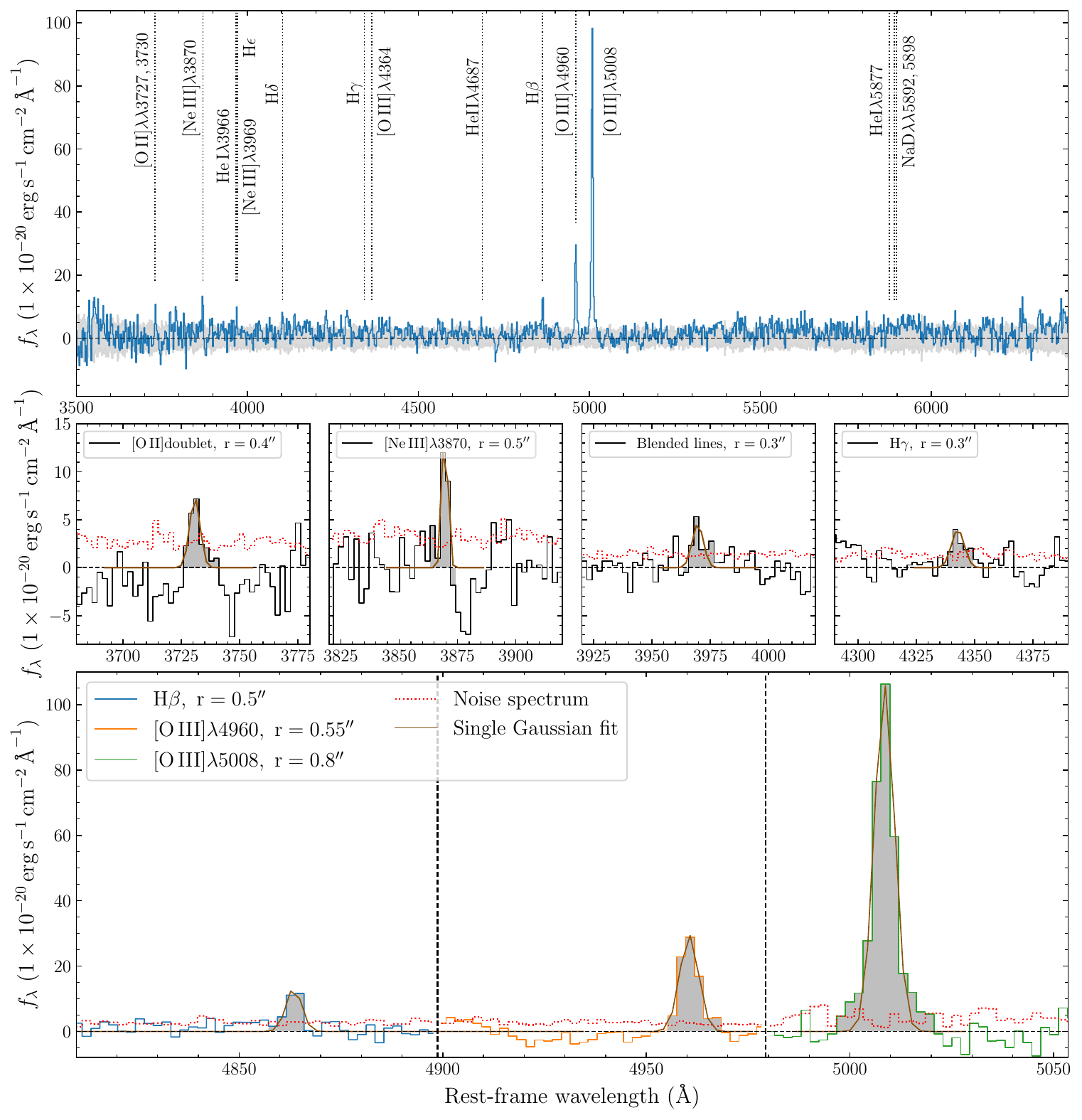}
    \caption{Spectra extracted from NIRSpec IFU data. Top panel: full spectrum obtained from a circular aperture with a radius of $0\farcs5$, with key emission and absorption lines labeled. The light grey shaded region indicates $1\,\sigma$ noise level. Middle and bottom panels: spectra of detected emission lines extracted using different aperture sizes to measure their total fluxes. Continuum levels were measured in corresponding aperture sizes and subtracted. From left to right, the bottom panel shows H$\beta$ and \oiii$\lambda$4960, 5008 lines, while the middle panel displays \oii$\lambda$3727, 3730, \neiii3870, \hei3966+\neiii3969+H$\epsilon$ blended and H$\gamma$ lines. The radii of circular apertures from which the spectra were extracted to measure total fluxes are shown in the labels. The red dotted line corresponds to the noise spectra measured from random aperture photometry using the same aperture sizes as the spectral extractions. Black vertical dashed lines demarcate the wavelength ranges for which the spectra were generated with different aperture sizes. Brown solid lines indicate single Gaussian fits to the emission lines, and grey shaded regions highlight integration ranges for each line.}
    \label{spec}
\end{figure*}

\begin{figure*}
    \includegraphics[width=\textwidth]{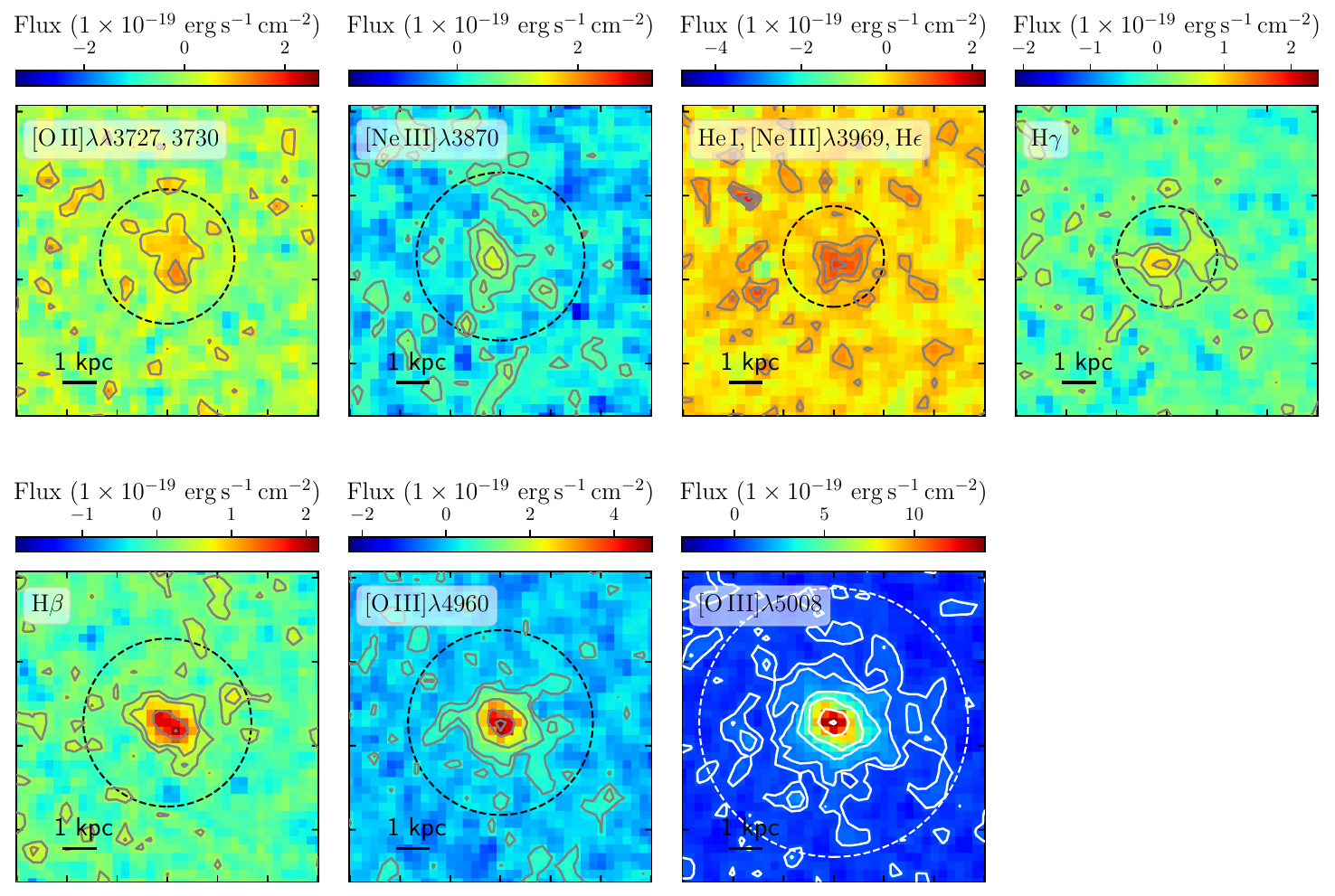}
    \caption{Moment 0 maps of detected lines using NIRSpec IFU data. The dashed black and white circles indicate the circular apertures from which we extracted the spectra to measure total fluxes. In upper panels, the grey contours indicate [1, 2, 3]$\sigma$ significance levels, where $1\,\sigma$ noise level is the standard deviation measured after $3\,\sigma$ clipping over the entire FoV on each moment-0 map, which is different from the method we used for measuring the uncertainties of line fluxes and actual S/Ns. Similarly, in bottom panels, the grey and white contours show [1, 2, 5, 7]$\sigma$, [1, 2, 5, 10, 13]$\sigma$ and [1, 2, 5, 10, 20, 28]$\sigma$ significance levels of H$\beta$ and [O\,{\sc iii}] doublet lines, respectively.}
    \label{moment 0 map}
\end{figure*}

On the top panel of Fig. \ref{spec}, we show the full 1D spectrum extracted from a circular aperture with a radius of $0\farcs5$ and centered at the peak position of \oiii5008 in NIRSpec IFS data. Though no significant continuum emission is detected above the noise level, the positive signals are continuously distributed across the spectrum. We therefore fitted the spectrum using a 1D polynomial model while masking out all potential emission line regions. The fitted model was then subtracted from the original spectrum to enable accurate measurements of line fluxes. After that, we conducted curve of growth analysis on the continuum-subtracted spectrum to identify detected emission lines and measure their total fluxes. Similar to NIRCam flux measurements, the curve of growth analysis allows us to measure the total flux experimentally, and no further aperture correction was applied. We detected \oii$\lambda$3727, 3730 with $r = 0\farcs4$, \neiii3870, \hei3966+\neiii3969+H$\epsilon$ blended lines and H$\gamma$ with $r = 0\farcs3$, H$\beta$ with $r = 0\farcs5$, \oiii4960 with $r = 0\farcs55$ and \oiii5008 with $r=0\farcs8$. The moment 0 maps of these lines are shown in Fig. \ref{moment 0 map}. The center of the apertures is the peak position of \oiii5008 emission. The aperture radius to measure the total flux, the measured line flux, S/N and FWHM of each line are summarized in Table \ref{line results}. We found no evidence of AGN activity because there is no clear detection of broad components of Balmer lines.

\begin{table}
    \centering
    \caption{Measurements of emission lines. Values in the top subtable are measured from single Gaussian fittings. Results in the bottom subtable are derived from double Gaussian fittings.}
    \label{line results}
    \begin{threeparttable}
    \begin{tabular}{lcccc}
    \hline
     & R\tnote{$^\ast$} & Flux & S/N\tnote{$^\intercal$} & FWHM\tnote{$^\dagger$} \\
     & (\arcsec) & ($10^{-18}\rm{erg}\,\rm{s}^{-1}\,\rm{cm}^{-2}$) & & $(\rm{km}\,\rm{s}^{-1})$ \\
    \hline
    [O\,{\sc ii}]3727,30 & 0.4 & $3.8\pm1.1$ & 3.6 & - \\
    \neiii3870 & 0.5 & $4.4\pm1.4$ & 3.1 & - \\
    Blended lines\tnote{$^{\ddagger}$} & 0.3 & $2.4\pm0.6$ & 4.0 & $220\pm203$ \\
    H$\gamma$ & 0.3 & $2.2\pm0.5$ & 4.0 & $298\pm107$ \\
    H$\beta$ & 0.5 & $5.2\pm1.3$ & 3.9 & - \\
    \oiii4960 & 0.55 & $14.5\pm1.5$ & 9.5 & $152\pm50$ \\
    \oiii5008 & 0.8 & $60.1\pm3.2$ & 18.6 & $193\pm30$ \\
    \hline
    \multicolumn{5}{c}{\oiii5008} \\
    \hline
    Narrow & 0.8 & $32.2\pm8.3$ & 3.9 & - \\
    Broad & 0.8 & $26.4\pm8.1$ & 3.3 & $629\pm162$ \\
    \hline
    \end{tabular}
    \begin{tablenotes}
    \item Note:
    \item[$^\ast$]Radius of the circular aperture to measure the total flux of each line.
    \item[$^\dagger$] Intrinsic FWHM measured by deconvolving the apparent FWHM with the instrumental resolution. For lines we do not show their intrinsic FWHMs, their observed FWHMs are not resolved by the instrumental resolutions at corresponding wavelengths.
    \item[$^\ddagger$] Blended \hei3966, \neiii3969 and H$\epsilon$ lines.
    \item[$^\intercal$] The method of measuring uncertainties of the line fluxes is outlined in \S\ref{Measure Line Fluxes and FWHMs}.
    \end{tablenotes}
    \end{threeparttable}
\end{table}

\subsection{Measuring Line Fluxes and FWHMs}
\label{Measure Line Fluxes and FWHMs}
To measure line fluxes, we integrated line profiles over spectral widths determined through visual inspection, summing the area under spectral profiles within the defined line regions. The integrated regions are wide enough because the edge bins are below or around $1\,\sigma$ level of adjacent noise, as shown by the grey shaded regions in bottom panels of Fig. \ref{spec}. To measure the uncertainty, we generated random apertures that do not overlap each other around the source and extracted spectra from them. We then calculated the standard deviation of the area under the random aperture spectra over the same spectral ranges when calculating flux. Notably, the line fluxes derived from line integrations and single Gaussian fittings are all compatible within $1\,\sigma$, except for \oiii5008 that exhibits broader wings in the line profile. We will discuss this broad component in the following section.

We measured the apparent FWHM of each line by performing single Gaussian fitting. We then measured the intrinsic FWHM by deconvolving the apparent FWHM with the instrumental spectral resolution at the corresponding wavelength. The intrinsic FWHMs of optical lines are listed in Table \ref{line results}. They are consistent with the FWHM of \oeight~emission ($152.9\pm9.0\,\rm{km}\,\rm{s}^{-1}$) measured in \citet{yi23} within $1-2\,\sigma$.

\subsection{Broad Component of \oiii$\lambda$4960, 5008}
\label{double gauss fit}
\begin{figure}
    \centering
    \includegraphics[width=\linewidth]{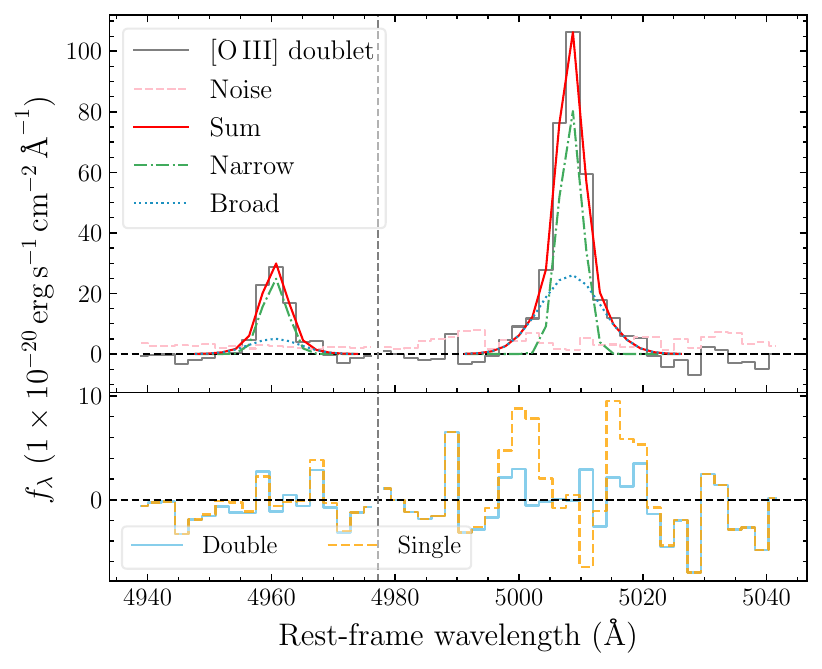}
    \caption{Double Gaussian fitting results and residual analysis for \oiii$\lambda$4960, 5008 lines. In the top panel, the solid grey line shows the spectrum extracted from the circular aperture to measure the total flux of each line. The grey vertical dashed line demarcates the wavelength ranges for which the spectra were generated with different aperture sizes. The dashed pink line shows the noise spectrum obtained from random aperture photometry with same aperture size to measure the total flux of the corresponding line. The dotted dashed green line shows the narrow component, the dotted blue line shows the broad component and the solid red line shows the sum of each line's narrow and broad components. The bottom panel shows the spectral residual after subtracting the single (orange dashed) or double (blue solid) Gaussian fitting curve.}
    \label{spec_dbgauss}
\end{figure}

As shown in Table \ref{line results}, the flux ratio of [O\,{\sc iii}] doublet is much larger than the theoretical value of 2.98 \citep{storey}. Furthermore, as can be seen from the bottom panel of Fig. \ref{spec}, the \oiii5008 line profile shows a broader component compared to the single Gaussian profile. Therefore, we performed double Gaussian fitting on \oiii5008 to measure the broad component. We also conducted residual analysis to examine its reliability. Specifically, we subtracted the single or double Gaussian fitting curves from the continuum-subtracted spectrum and then compared the significance of residuals. In the double Gaussian fitting in this section, we set the central wavelengths of the narrow and broad components to be the same. The measurements of narrow and broad \oiii5008 are listed in Table \ref{line results}. The spectral residual after subtracting single and double Gaussian fittings is shown in Fig. \ref{spec_dbgauss}. The S/Ns of narrow and broad components of \oiii5008 are $>3\,\sigma$, and the spectral residual after subtracting double Gaussian fitting is much smaller than that after single Gaussian fitting; thus, the detection of the broad \oiii5008 component is robust. The intrinsic FWHM of broad \oiii5008 is $629\pm162\,\rm{km}\,\rm{s}^{-1}$, indicating the existence of ionized gas outflows. We will discuss the outflows in detail in \S\ref{outflow}.

Since the \oiii$\lambda$4960, 5008 lines should originate from the same regions and have similar properties, the \oiii4960 line should also have a broad component. Therefore, we performed double Gaussian fitting and residual analysis on \oiii4960 as well. Consequently, although the broad \oiii4960 component is too weak to claim its detection, the flux ratio of narrow \oiii5008 and narrow \oiii4960 ($\sim3.1$) is consistent with the theoretical ratio of $\sim3$. Deeper integrations would be required to establish the presence of the broad line component in \oiii4960.

\subsection{Measure Metallicity From Spectral Lines}
\label{strong line method}
We employed strong-line method to estimate metallicity using the following diagnostics: R3 (\oiii5008/H$\beta$), R2 (\oii$\lambda$3727, 3730/H$\beta$), R23 ((\oii$\lambda$3727, 3730 + \oiii$\lambda$4960, 5008)/H$\beta$) and O3O2 (\oiii5008/\oii$\lambda$3727, 3730) adopted in \citet{curti20}. 

These diagnostics are based on line ratios corrected for dust extinction using the Milky Way extinction law from \citet{cardelli}. For our target, we determined the dust attenuation through SED fitting, and the derived value is $A_{\rm{V}}=0.21\pm0.06$ mag (Table \ref{SED results}). Although we adopted \citet{calzetti} starburst reddening curve in the SED fitting, the differences between various dust attenuation curves in rest-optical range are minimal (e.g., \citealt{Battisti22}). Therefore, we applied \citet{calzetti} law to correct for dust extinction of emission lines. Though \citet{calzetti} suggests $E(B-V)_{\rm{stellar}} = 0.44\,E(B-V)_{\rm{nebular}}$, considering that we adopted a multiplicative factor of one in the SED fitting and the stellar age of our target is very young (\S\ref{sed}), no additional correction for nebular reddening was applied to the emission lines. 

The marginally detected H$\gamma$ emission enables us to measure dust attenuation through Balmer decrement. Since H$\gamma$ is detected with slightly smaller aperture size compared to most of other lines, the dust attenuation inferred from H$\gamma$/H$\beta$ ratio constrains the physical conditions of more central area of the galaxy. The observed H$\gamma$/H$\beta$ ratio is $0.459\pm0.122$, when the H$\beta$ flux is measured from the same aperture size as H$\gamma$. Adopting an intrinsic H$\gamma$/H$\beta$ ratio of 0.468 for case B recombination at $T_{\rm{e}} = 10,000\,\rm{K}$ and $n_{\rm{e}} = 100\,\rm{cm}^{-3}$ \citep{osterbrock}, the derived dust attenuation is $A_{\rm{V}} \sim 0.16$, consistent with the values obtained from SED fittings well. When considering a range of $T_{\rm{e}}$ from 5,000 to 20,000 K, the measured ratio is compatible with no extinction (5,000 K) to $A_{\rm{V}} \sim 0.28$ (20,000 K). The measured line ratio also has large uncertainties. Therefore, we adopted $A_{\rm{V}}$ obtained from SED fitting for dust extinction correction throughout this paper.

The values of $\log \rm{R}3, \log \rm{R}23~\rm{and} \log \rm{O3O2}$ calculated with the total flux of \oiii5008 exceed the maximum values of the model fits for calibration of the diagnostics. To address the saturation of R3 and R23, we determined the metallicities as the modeled metallicities at the maximum values of R3 and R23. Considering the dispersion of the calibration,  this approach resulted in $12+\log(\rm{O/H})_{\rm{R3}} = 8.00\pm0.07$ and $12+\log(\rm{O/H})_{\rm{R23}} = 8.07\pm0.12$. Using only flux of narrow [O\,{\sc iii}] does not change the results for R3 and R23, while we obtained $12+\log(\rm{O/H})_{\rm{O3O2}} = 7.79\pm0.14$, slightly lower than the above values. On the other hand, $\log\rm{R2}$ yields a metallicity of $12+\log(\rm{O/H})_{\rm{R2}} = 7.82\pm0.20$. These measurements are consistent with each other within uncertainties. We take the metallicity derived from R23 diagnostics as the fiducial value from strong-line method. $12+\log(\rm{O/H})_{\rm{R23}}$ corresponds to $Z/\rm{Z}_\odot = 0.24\pm0.07$, consistent with the metallicity derived from SED fitting in \S\ref{sed}.

\subsection{Measure SFR From H$\beta$ Line}
\label{sfr from hb}
We measured SFR based on H$\alpha$ emission inferred from the observed H$\beta$ emission. Dust extinction was corrected following the same procedure described in \S\ref{strong line method}. We assumed the intrinsic intensity ratio of H$\alpha$ and H$\beta$ is 2.863 for case B recombination at $T_{\rm{e}} = 10,000\,\rm{K}$, $n_{\rm{e}} = 100\,\rm{cm}^{-3}$ \citep{osterbrock}. We then obtained the intrinsic luminosity of H$\alpha$ emission and applied the conversion factor for H$\alpha$ from \citet{kennicutt2012} to infer the SFR. The resulting value is $\log(\rm{SFR}/\rm{M}_\odot\,\rm{yr}^{-1}) = 1.78\pm0.12$. Since the calibration of \citet{kennicutt2012} assumes solar metallicity, and our target has a significantly lower metallicity, we corrected for the metallicity effect using the spectral model from \citet{inoue11}. Specifically, \citet{inoue11} shows that, for a constant star formation of 10 Myr and a LyC escape fraction of $f_{\rm{esc}} = 0$, the ratio of H$\beta$ luminosity and SFR at $Z\sim0.2\,\rm{Z}_\odot$ is approximately $16/11$ times larger than the ratio at $Z\sim \rm{Z}_\odot$. Consequently, the SFR corrected for the metallicity effect is reduced by a factor of $\sim11/16$, yielding $\log(\rm{SFR}/\rm{M}_\odot\,\rm{yr}^{-1}) = 1.62\pm0.12$. The assumption of electron temperature has a negligible effect on the final results. For instance, supposing $T_{\rm{e}} = 20,000\,\rm{K}$ only reduces the estimated SFR by 4 per cent. 

Aside from metallicity, the derived SFR is also very sensitive to stellar age when it is $\lesssim10$ Myr. \citet{sugahara22} established relation between H$\beta$ luminosity and SFR considering effects from simple stellar population (SSP) models, metallicity and stellar ages. Adopting Starburst99 models \citep{leitherer99, leitherer14}, also used for calibrations in \citet{kennicutt2012}, a metallicity of $Z\sim0.2\,\rm{Z_\odot}$ and a stellar age of 10 Myr yields the same SFR as derived above. However, adopting an age of 1.8 Myr (i.e. the mass-weighted mean age) increases the obtained SFR by 0.3 dex, resulting in $\log(\rm{SFR}/\rm{M}_\odot\,\rm{yr}^{-1}) = 1.93\pm0.12$. Assuming BPASS models which incorporate massive star binaries and better represent high-$z$ galaxies \citep{eldridge17} does not result in a very different SFR, which decreases by only 0.08 dex and 0.05 dex for age of 10 Myr and 1.8 Myr, respectively. 

Previously, \citet{inoue16} measured the SFR using SED fitting with data from Subaru, UKIRT, and Spitzer telescopes. The derived SFR is $347^{+166}_{-279}\,\rm{M}_\odot\,\rm{yr}^{-1}$ with a stellar age of only $\sim1\,\rm{Myr}$. In this work, our SED fitting yields an instantaneous SFR of $\sim165\,\rm{M}_\odot\,\rm{yr}^{-1}$ and a mass-weighted mean age of $\sim2\,\rm{Myr}$ (Table \ref{SED results}). Given that the lifetimes of stars traced by emission lines from ionized gas are typically $\sim3-10$ Myr \citep{kennicutt2012}, averaging the instantaneous SFR over 10 Myr would yield a SFR of $\sim1/5\times\rm{SFR}_{\rm{SED}}=33\,\rm{M}_\odot\,\rm{yr}^{-1}$, consistent with the SFR derived from H$\beta$. 

\section{Discussion}
\subsection{Optical [O\,{\sc iii}] Outflows}
\label{outflow}
As discussed in \S\ref{double gauss fit}, we identified a broad component in the \oiii5008 emission, indicative presence of ionized gas outflows. In this section, we present analysis and discussions about the properties of the outflows.

\subsubsection{Outflow Velocity}
\label{outflow velocity}
 We derived the outflow velocity using equation (1) in \citet{carniani24}. Unlike the approach in \S\ref{double gauss fit}, we performed double Gaussian fitting on \oiii5008 without constraining the central wavelengths of narrow and broad components to be identical in order to measure the velocity offset between them. The resulting velocity offset is $-0.7\pm43.7\,\rm{km}\,\rm{s}^{-1}$. We then substituted the velocity offset and the intrinsic velocity dispersion of the broad component to the equation, obtaining an outflow velocity of $535\pm132\,\rm{km}\,\rm{s}^{-1}$. The uncertainty is measured by propagating the uncertainties of both the velocity offset and the velocity dispersion. This velocity is consistent with $z\sim3.5-6.9$ JADES samples \citep{carniani24}. It also agrees with the outflow velocities measured from metal absorption lines in $z\sim5-6$ SFGs \citep{sugahara19}.

 We further estimated the escape velocity to determine if the outflowing material can escape from the gravitational potential of the dark matter halos of our target. We estimated the halo mass based on the stellar mass-halo mass relation at $z=7$ from \citet{behroozi13}, yielding $M_{\rm{h}}\sim5\times10^{10}\,\rm{M}_\odot$. We then derived the virial radius using the following equation:
 \begin{equation}
     r_{\rm{vir}} = \left(\frac{3M_{\rm{h}}}{4\pi200\rho_{\rm{crit}}\left(z\right)}\right)^{1/3},
 \end{equation}
 where $\rho_{\rm{crit}}\left(z\right)$ is the critical density of the Universe at a given redshift. The resulting virial radius is 13.8\,kpc. Furthermore, adopting the following equation:
 \begin{equation}
     v_{\rm{esc}} = \left(\frac{2GM_{\rm{h}}}{r_{\rm{vir}}}\right)^{1/2},
 \end{equation}
 we estimated an escape velocity of $\sim177\,\rm{km}\,\rm{s}^{-1}$, approximately three times smaller than the outflow velocity. However, the outflow velocity should be treated as an upper limit according to its definition, and only a small fraction of the outflowing material is moving with this high velocity. Thus, only part of the outflows may escape the gravitational potential of the dark matter halos and it might have negligible effect on quenching the galaxy.
 
\subsubsection{Mass Loading Factor}
\label{mass loading factor}
\begin{figure}
    \centering
    \includegraphics[width=\linewidth]{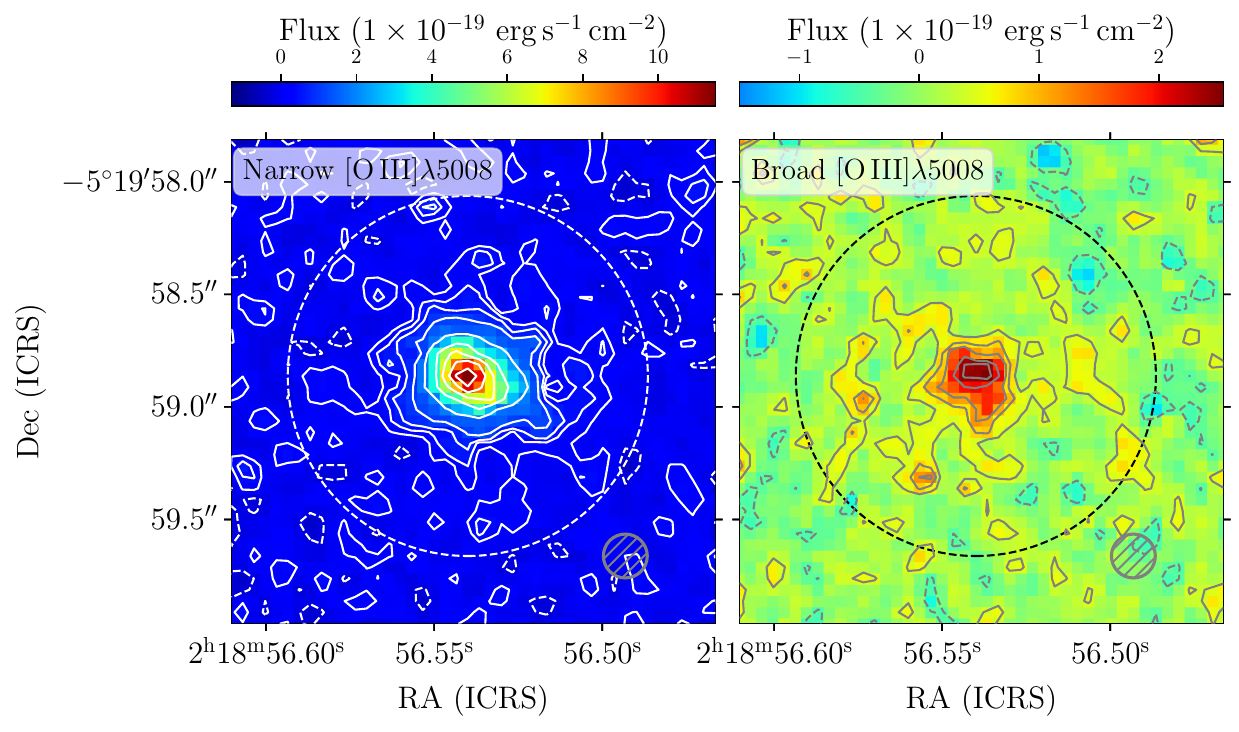}
    \caption{Left panel: moment-0 map of narrow component of \oiii5008 emission. The integral range is from $-200$ to $+200\,\rm{km}\,\rm{s}^{-1}$. The white dashed and solid lines are [-1, 1, 2, 3, 5, 10, 20, 30, 40, 50] $\sigma$ contours of narrow \oiii5008 emission. Right panel: moment-0 map of broad component of \oiii5008 emission. The integral range is from $-720$ to $-200\,\rm{km}\,\rm{s}^{-1}$ and from $+200$ to $+850\,\rm{km}\,\rm{s}^{-1}$. The grey dashed and solid lines are [-1, 1, 2, 3, 5, 6] $\sigma$ contours of broad \oiii5008 emission. Each panel's $1\,\sigma$ noise level is the standard deviation measured after $3\,\sigma$ clipping over the entire FoV on each moment-0 map. The white dashed circle in the left panel and the black dashed circle in the right panel indicate the aperture to measure the total flux of \oiii5008 emission. The grey hatched circle in each panel shows the PSF FWHM of NIRSpec spectrum at the observed wavelength of \oiii5008.}
    \label{na_br_5007}
\end{figure}

We constructed moment-0 maps of narrow and broad \oiii5008 emission by integrating over the velocity ranges corresponding to the narrow and broad components, respectively (Fig. \ref{na_br_5007}). We employed the following equation to calculate the mass outflow rate:
\begin{equation}
    \dot{M}_{\rm{out}} = v_{\rm{out}}M_{\rm{out}}/R_{\rm{out}}
\label{equation 1}
\end{equation}
where $v_{\rm{out}}$ is the outflow velocity, $M_{\rm{out}}$ is the outflow mass, and $R_{\rm{out}}$ is the radius of the outflow region. We adopted the half-light radius of the broad \oiii5008 component as the outflow radius. We determined this radius through curve of growth analysis on moment-0 map of the broad \oiii5008 using Photutils \citep{bradley24}. We added a 31 per cent uncertainty derived from line integration for broad \oiii5008 (Table \ref{line}) to the total flux of broad [O\,{\sc iii}], yielding a value of $r_{\rm{out}} = 1.86\pm0.63$ kpc (Table \ref{many result}). We did not account for the PSF size ($\sim1$ kpc) in this estimation, because it has a negligible impact on the derived size and mass outflow rate ($\sim0.02$ dex). The outflows would require $\sim3.4$ Myr to reach this spatial extent, given the measured outflow velocity. 
Notably, the outflow radius of our galaxy is approximately a factor of 4.8 larger than the median value of the JADES samples, as they were determined from half-light radius of galaxies in NIRCam images \citep{carniani24}. Similarly, for our target, the half-light radii of UV continuum in SW images and optical emission in F444W are $\sim0.46$ and $\sim0.78$ kpc, approximately a factor of 4 and 2.4 smaller than the one derived from broad \oiii5008, respectively.

\begin{table}
    \centering
    \caption{Summary of physical properties aside from direct outputs from SED fitting. Uncertainty is the $1\,\sigma$ deviation.}
    \label{many result}
    \begin{threeparttable}
    \renewcommand{\arraystretch}{1.2}
    \begin{tabular}{lc}
    \hline
    Property & Value \\
    \hline
    UV slope & $-2.32\pm0.04$ \\
    $(12+\log\rm{(O/H)})_{R23}$ & $8.07\pm0.12$ \\
    SFR$_{\rm{H\beta}}$ $(\log (\rm{M}_\odot\,\rm{yr}^{-1}))$ & $1.62\pm0.12$ \\
    SFR$_{10}$ $(\log (\rm{M}_\odot\,\rm{yr}^{-1}))^{\dagger}$ & $1.58\pm0.05$ \\
    $r_{\rm{e}}^{\rm{[OIII],broad}}$ (kpc) & $1.86\pm0.63$ \\
    $r_{\rm{e}}^{\rm{[OIII],narrow}}$ (kpc) & $0.90\pm0.23$ \\
    $M_{\rm{UV}}$ (mag) & $-22.1\pm0.2$ \\
    EW$_{\rm{Ly\alpha}}$~(\r{A}) & $30^{+39}_{-14}$ \\
    EW$_{\rm{[OIII]+H\beta}}$~(\AA) & $3700\pm370$ \\
    $\log(Q/\rm{s}^{-1})$ &  $54.9\pm0.1$ \\
    $\log(\xi_{\rm{ion}}/\rm{erg}^{-1}\,\rm{Hz})$ & $25.44\pm0.15$ \\
    $\Sigma_{\rm{SFR,SED}}~(\rm{M}_\odot\,\rm{yr}^{-1}\,\rm{kpc}^{-2})$\tnote{$^\ast$} & $32.6\pm17.2$ \\
    $\Sigma_{\rm{SFR,H\beta}}~(\rm{M}_\odot\,\rm{yr}^{-1}\,\rm{kpc}^{-2})$\tnote{$^\ddagger$} & $8.2\pm4.8$ \\
    $t_{\rm{depl,SED}}$ (Myr)\tnote{$^\ast$} & $114^{+57}_{-42}$ \\
    $t_{\rm{depl,H\beta}}$ (Myr)\tnote{$^\ddagger$} & $445^{+273}_{-178}$ \\
    \hline
    \multicolumn{2}{c}{\rm{Outflow Properties}} \\
    \hline
    $v_{\rm{out}}$ $(\rm{km}\,\rm{s}^{-1})$ & $535\pm132$ \\
    $M_{\rm{out}}$ ($\log\rm{M}_\odot$) & $8.02\pm0.43$ \\
    $\dot{M}_{\rm{out}}$ ($\log (\rm{M}_\odot\,\rm{yr}^{-1})$) & $1.49\pm0.47$ \\
    $\log \eta$\tnote{$^\ast$} & $-0.73\pm0.47$ \\
    $\log \eta_{10}$\tnote{$^\dagger$} & $-0.09\pm0.47$ \\
    \hline
    \multicolumn{2}{c}{\rm{Mass Properties} ($\times10^{10}\,\rm{M}_\odot)$} \\
    \hline
    $M_{\rm{dyn}}^{\rm{disp}}$ & $1.71\pm1.57$ \\
    $M_{\rm{dyn}}^{\rm{rot}}$ & $7.13\pm6.54$ \\
    $M_{\rm{mol}}$ & $0.46^{+0.33}_{-0.23}$ \\
    $M_{\rm{HI}}$ & $1.42^{+0.84}_{-0.61}$ \\
    $M_{\rm{gas}}$ & $1.93^{+0.89}_{-0.68}$ \\
    $M_{\rm{gas}}+M_\ast$ & $1.97^{+0.89}_{-0.68}$ \\
    Potential old mass\tnote{$^\intercal$} ($\log(M_{\ast}^{\rm{old}}/\rm{M}_\odot)$) & $<9.5$ \\
    \hline
    \end{tabular}
    \begin{tablenotes}
    \item Note:
    \item[$^\ast$]Derived using instantaneous SFR from SED fitting.
    \item[$^\ddagger$]Determined using H$\beta$-based SFR.
    \item[$^\dagger$]Averaged SFR over 10 Myr, or derived from 10 Myr averaged SFR.
    \item[$^\intercal$]Defined as the stellar mass formed when a bursty star formation happens at $z = 20$.
    \end{tablenotes}
    \end{threeparttable}
\end{table}

The outflow mass was determined using the equation (6) in \citet{carniani24}:
\begin{equation}
    M_{\rm{out}} = 0.8\times10^8\left(\frac{L_{\rm{[OIII]}}^{\rm{broad,corr}}}{10^{44}\,\rm{erg}\,\rm{s}^{-1}}\right)\left(\frac{Z_{\rm{out}}}{Z_\odot}\right)^{-1}\left(\frac{n_{\rm{out}}}{500\,\rm{cm}^{-3}}\right)^{-1}\,\rm{M}_\odot.
\end{equation}
We assumed the outflow metallicity to be the gas-phase metallicity derived from SED fitting (Table \ref{SED results}), following \citet{carniani24}. The predicted electron density for galaxies at $z=7.212$ is $\log (n_e/\rm{cm}^{-3}) = 2.83\pm0.43$ \citep{abdurro24}, while the electron density of our target constrained by \ofive~emission has a $1\,\sigma$ upper limit of approximately $<150\,\rm{cm}^{-3}$ (Ren at al. in preparation). The densities in \citet{abdurro24} were derived through [O\,{\sc ii}] doublet which may trace different ionized gas regions compared to [O\,{\sc iii}], leading to the density variations in the two measurements. Similarly, \citet{carniani24} adopted a typical electron density of $380\,\rm{cm}^{-3}$ in outflows at $z\sim2$, determined by [S\,{\sc ii}] doublet \citep{foster19}. Consequently, we adopted the typical value of $380\,\rm{cm}^{-3}$ and assumed a 0.4 dex uncertainty, considering both the constraints from \ofive~emission and the density-redshift evolution. Additionally, we corrected for dust extinction in broad \oiii5008 following the same procedure outlined in \S\ref{strong line method}. The derived outflow mass is $\log (M_{\rm{out}}/\rm{M}_\odot) = 8.02\pm0.43$, resulting in a mass outflow rate of $\log (\dot{M}_{\rm{out}}/\rm{M}_\odot\,\rm{yr}^{-1}) = 1.49\pm0.47$. The $M_{\rm{out}}$ and $\dot{M}_{\rm{out}}$ of our target are 1.4 and 1.0 dex higher than the median values of the JADES samples, respectively. These differences likely arise from the bright [O\,{\sc iii}] broad component of our target, which has a peak flux density up to $\sim2$ orders of magnitude higher than that of JADES samples \citep{carniani24}.

Using the instantaneous SFR measured from SED fitting and adopting the relation $\eta = \dot{M}_{\rm{out}}/\rm{SFR}$, the derived mass loading factor is $\log \eta = -0.73\pm0.47$. The median of this value is lower than that of all the JADES samples and the prediction by simulations. However, it aligns with the observed anti-correlation between $\eta$ and $M_\ast$, given that the stellar mass of our target is comparable to the upper limit of the stellar mass range of the samples. This low value also results from the bursty SF phase in our galaxy, where the instantaneous SFR reaches $\sim165\,\rm{M}_\odot\,\rm{yr}^{-1}$, compared to the 10 Myr-averaged SFRs adopted in \citet{carniani24}. To facilitate a direct comparison, we estimated the SFR averaged over 10 Myr for our galaxy by dividing the stellar mass by 10 Myr, resulting in a value of $\log(\rm{SFR}_{10}/\rm{M}_\odot\,\rm{yr}^{-1}) = 1.58\pm0.05$. We note that this SFR agrees well with the SFR derived from H$\beta$ emission (Table \ref{many result}; \S\ref{sfr from hb}). While it is still 1.2 dex higher than the median SFRs of JADES samples, the adjusted mass loading factor of $\log\eta = -0.09\pm0.47$ is consistent with both the JADES samples and the predicted value for ionized outflows from FIRE-2 simulation at the stellar mass of our target (\citealt{pandya21, carniani24}). Notably, since the outflow velocity represents the highest speed of the outflows, the derived mass loading factor may also represent an upper limit.

\citet{fluetsch19} studied multiphase outflows in local star-forming galaxies and AGNs, finding that starburst-driven ionized outflows and molecular outflows exhibits similar mass outflow rates. The comparison of neutral atomic outflows and molecular outflows is more complex. While their mass outflow rates typically differ by less than one order of magnitude, the sample volume is very limited. Consequently, the total mass outflow rate and mass loading factor of our target could be up to $\sim1$ dex higher than the measurements of ionized outflows alone. So far, ALMA observations have found no broad component of \ctwo~emission in our target (\citealt{inoue16, ca20, yi23}). However, given that [C\,{\sc ii}] was marginally detected at $\sim4\,\sigma$, deeper observations are necessary to identify the potential broad component and probe neutral outflows.

We attempt to estimate the properties of neutral gas outflows using sensitivity of ALMA [C\,{\sc ii}] dataset. The outflows traced by broad [C\,{\sc ii}] are mostly in the neutral atomic phase because the majority of the [C\,{\sc ii}] emission comes from gas in the atomic phase. Assuming that the potential broad component of [C\,{\sc ii}] has the same FWHM as broad \oiii5008, we generated a moment-0 map of the potential broad [C\,{\sc ii}] by integrating over -630 to +620 km$\,\rm{s}^{-1}$ (approximately twice the FWHM of broad [O\,{\sc iii}]) while excluding the channels that cover the narrow [C\,{\sc ii}] \citep[-80 to +145 km$\,\rm{s}^{-1}$][]{yi23}. We measured the $3\,\sigma$ upper limit for the broad [C\,{\sc ii}] luminosity, which is $L^{\rm{broad}}_{\rm{[CII]}}<1.0\times10^8\,\rm{L}_\odot$. Then, to estimate the atomic outflow mass from the broad [C\,{\sc ii}] luminosity, we adopted the conversion factor $\kappa_{[\rm{CII}]}$ given by equation (C.3) in \citet{Herrera-Camus21}. Specifically, we assumed a neutral gas temperature of $T=100\,$K to derive the critical density for [C\,{\sc ii}] against hydrogen atoms, which is $\sim3\times10^3\,\rm{cm}^{-3}$. We adopted the metallicity measured from R23 diagnostics to estimate the carbon abundance, yielding $\chi_{C^+} = 1.9\times10^{-5}$. We also assumed a neutral gas density of $n=10^4\,\rm{cm}^{-3}$. Therefore, we obtained a conversion factor of $\kappa_{[\rm{CII}]} = 20.2\,\rm{M}_\odot\rm{L}_\odot^{-1}$, resulting in an atomic outflow mass of $<2.0\times10^9\,\rm{M}_\odot$. Assuming that the atomic gas outflows has the same velocity and spatial extent as the ionized outflows, the mass outflow rate and mass loading factor of atomic gas outflows are $\log(\dot{M}_{\rm{HI,~out}}/\rm{M}_\odot\,\rm{yr}^{-1}) < 2.8$ and $\log(\eta_{\rm{10,~HI}}) < 1.2$, respectively. These values are $<1.3$ dex larger than those of ionized gas outflows. Recent observations of ionized and atomic gas outflows in a main-sequence SFG at $z\sim5.5$ reported an atomic mass outflow rate more than one dex larger than the ionized one \citep{parlanti25}. Considering that our estimates are upper limits, our results may be consistent with recent high-$z$ observations. However, we have to note that these estimates are very uncertain because they are derived based on many assumptions.

The total mass loading factor of all phases of gas may be larger than one, in which case the outflow-driven mechanisms are likely to quench the star formation effectively \citep[e.g.,][]{carniani24}. However, considering the large uncertainties of the estimated mass loading factors, it is uncertain whether the outflows can effectively quench the star formation in our target.

\subsection{Dynamical Mass, Gas Mass and Gas Depletion Time}
\subsubsection{Dynamical Mass}
\label{dynamical mass}
We estimated the dynamical mass of our target using ALMA observations of \ctwo~emission \citep{yi23}. The [C\,{\sc ii}] observations yield a detection of $4.5\,\sigma$ or $3.6\,\sigma$ when integrating with or without one suspicious edge bin. Our analysis on both dynamical mass and gas mass primarily adopts the $4.5\,\sigma$ detection but includes discussion of potential effects arising from the $3.6\,\sigma$ detection. Given the absence of spatially resolved velocity information, we computed dynamical mass under both dispersion- and rotation-dominated assumptions. For the dispersion-dominated scenario, we applied equation (6) from \citet{decarli18}:
\begin{equation}
    M_{\rm{dyn}} = \frac{3}{2}\frac{\it{R}_{\rm{[CII]}}\sigma^2_{\rm{line}}}{\it{G}}.
\end{equation}
The radius of the [C\,{\sc ii}]-emitting region was taken as the beam-deconvolved semi-major axis FWHM of the source size ($5.1\pm2.7$ kpc), derived from 2D Gaussian fitting. However, the [C\,{\sc ii}] distribution is elongated with the minor axis being unresolved, leading to substantial uncertainty in the size measurement (see Fig. 6 in \citealt{yi23}). Additionally, we adopted a velocity dispersion of $98.1\pm36.9\,\rm{km}\,\rm{s}^{-1}$ determined by line FWHM ($231\pm87\,\rm{km}\,\rm{s}^{-1}$), which yielded a dynamical mass of $M_{\rm{dyn}} = (1.71\pm1.57)\times10^{10}\,\rm{M}_\odot$. The extremely large uncertainty ($\sim90$ per cent) is dominated by the uncertainty in velocity dispersion (38 per cent), which appears as a squared term in the equation.

In the rotation-dominated case, we assumed an inclination angle of $45\degr$ and employed equation (7) from \citet{decarli18}:
\begin{equation}
    M_{\rm{dyn}} = G^{-1}R_{\rm{[CII]}}\left(0.75\,\rm{FWHM}/sin{\it{i}}\right)^2,
\end{equation}
resulting in $M_{\rm{dyn}} = (7.13\pm6.54)\times10^{10}\,\rm{M}_\odot$. Since we assumed an average angle of $45\degr$, smaller than the assumption adopted in \citet{decarli18}, the resulting dynamical mass is approximately a factor of 4.2 larger than the dispersion-dominated scenario. The extremely large uncertainty arises from the same reason as the dispersion-dominated case. For the $3.6\,\sigma$ detection case, both major and minor axes of its distribution are unresolved. Assuming the major axis is roughly half that of the $4.5\,\sigma$ case based on visual inspection and the change of FWHM is minimal, the dynamical mass decreases by approximately half in both scenarios.

We also estimated dynamical mass using \oiii5008 emission. Considering that the outflowing component is not gravitationally bound to the galaxy, we derived the dynamical mass from the narrow [O\,{\sc iii}] component. The half-light radius of narrow [O\,{\sc iii}] (right panel of Fig. \ref{na_br_5007}) is determined through curve of growth analysis using Photutils \citep{bradley24}. Similar to the measurement of broad component (\S\ref{mass loading factor}), we propagated a 26 per cent uncertainty obtained from line integration for narrow \oiii5008 to the total flux of narrow [O\,{\sc iii}], yielding a value of $0.90\pm0.23$ kpc (Table \ref{many result}). Given that the line width of narrow [O\,{\sc iii}] is unresolved, we adopted the intrinsic [O\,{\sc iii}] line width derived from single Gaussian fitting (Table \ref{line results}). The obtained dynamical mass is $(2.1\pm0.8)\times10^9\,\rm{M}_\odot$ and $(8.8\pm3.5)\times10^9\,\rm{M}_\odot$ for dispersion- and rotation-dominated scenarios, respectively (deconvolving the radius with the PSF size would reduce the estimated mass by 0.8 dex). In both cases, the [O\,{\sc iii}]-based dynamical mass is one order of magnitude lower than the [C\,{\sc ii}]-based values, primarily because the [C\,{\sc ii}] radius is roughly a factor of 6 larger than that of narrow [O\,{\sc iii}]. This is expected, because [O\,{\sc iii}] originates from H\,{\sc ii} regions, while [C\,{\sc ii}] is predominantly emitted from more extended neutral gas clouds surrounding the ionized bubbles. ALMA observations of [O\,{\sc iii}] and [C\,{\sc ii}] emitters at $z>6$ have shown that the [C\,{\sc ii}] halos are typically more extended compared to [O\,{\sc iii}] (\citealt{ca20, fujimoto24_z8.5}).

\subsubsection{Gas Mass}
We further estimated molecular gas mass using the conversion factor between [C\,{\sc ii}] luminosity and molecular gas mass for high-$z$ galaxies, which is $\alpha_{\rm{[CII]}} = 35\,\rm{M}_\odot/L_\odot$ with a dispersion of 0.2 dex \citep{zanella18}. The [C\,{\sc ii}] luminosity for the $4.5\,\sigma$ detection is $(1.2\pm0.5)\times10^8\,\rm{L}_\odot$ \citep{yi23}, yielding a molecular gas mass of $M_{\rm{mol}} = 0.46^{+0.33}_{-0.23}\times10^{10}\,\rm{M}_\odot$. The median values and uncertainties were determined through the 16th, 50th and 84th percentiles of Monte Carlo simulations, with this approach applied for all relevant measurements. Additionally, we measured H\,{\sc i} gas mass using [C\,{\sc ii}]-to-H\,{\sc i} conversion relation from \citet{heintz21, Heintz_z8}. Adopting the metallicity derived from SED fitting and the [C\,{\sc ii}] luminosity of the $4.5\,\sigma$ detection, we obtained an H\,{\sc i} gas mass of $M_{\rm{HI}} = 1.42^{+0.84}_{-0.61}\times10^{10}\,\rm{M}_\odot$. Therefore, the total gas mass is $M_{\rm{gas}} = M_{\rm{mol}} + M_{\rm{HI}} = 1.93^{+0.89}_{-0.68}\times10^{10}\,\rm{M}_\odot$. Combined with the stellar mass determined from SED fitting, the total baryonic mass is $M_{\rm{gas}} + M_\ast = 1.97^{+0.89}_{-0.68}\times10^{10}\,\rm{M}_\odot$, consistent with the dynamical mass estimated for dispersion-dominated scenario. For case of [C\,{\sc ii}] detected at $3.6\,\sigma$, since the luminosity decreases by a factor of two \citep{yi23}, the inferred gas mass measurements similarly decrease by roughly half.

\subsubsection{Gas Mass Excess, Gas Mass Fraction and Gas Depletion Time}
From the mass measurements of different baryonic components, we derived a gas mass excess of $\log (M_{\rm{gas}}/M_\ast) = 1.70^{+0.17}_{-0.2}$ and a gas mass fraction of $f_{\rm{gas}} = M_{\rm{gas}}/(M_{\rm{gas}} + M_\ast) \sim98$ per cent. Assuming a constant star formation history with the instantaneous SFR derived from SED fitting, the gas depletion time is $t_{\rm{depl}} = 114^{+57}_{-42}$~Myr, defined as $M_{\rm{gas}}/\rm{SFR}$ without considering gas inflows and outflows. Alternatively, adopting the SFR measured from H$\beta$ emission yielded a longer depletion time of $t_{\rm{depl}} = 445^{+273}_{-178}$~Myr. These values of $t_{\rm{depl}}$ are consistent with simulated starbursts at high redshifts \citep{Ceverino18, dome24}. Different SFR tracers suggest our galaxy may be quenched at $z\sim6.5$ or $z\sim5$, respectively. With a current gas mass of $\sim1.9\times10^{10}\,\rm{M}_\odot$, the total stellar mass formed prior to quenching may reach a similar order of magnitude. Consequently, SXDF may be one of the progenitors of massive quiescent galaxies observed at $z\sim4-5$ (\citealt{carnall23, deGraaff25}).

For the case of $3.6\,\sigma$ [C\,{\sc ii}] detection, while the gas mass excess and gas depletion times decrease by a factor of $\sim2$, the gas mass fraction remains high at $\sim96\%$. \citet{Heintz_z8} reported that $z\sim4-8$ galaxies have elevated gas mass excess and shorter gas depletion times compared to $z\sim0$ samples. Our results align with the high-$z$ samples for both $4.5\,\sigma$ and $3.6\,\sigma$ cases. The extremely high gas mass fraction of our target is also consistent with some of the $z\gtrsim4$ samples \citep{Heintz_z8}.

\subsection{Spatial Distribution of UV continuum, Optical [O\,{\sc iii}] and \oeight~lines}
\begin{figure*}
    \includegraphics[width=\textwidth]{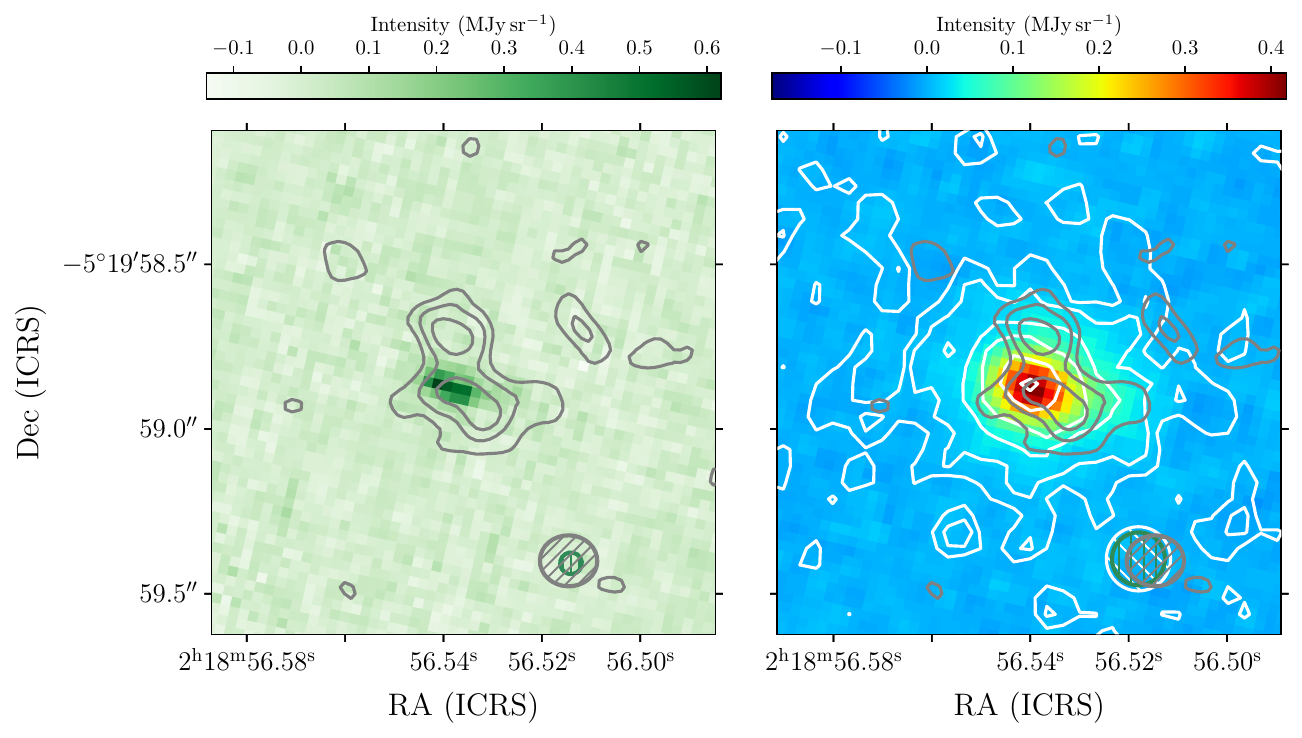}
    \caption{Left panel: Comparison of spatial distribution of UV continuum and \oeight~emission. Right panel: Comparison of optical and FIR [O\,{\sc iii}] emission. The background images are from F115W in the left panel and from F444W in the right panel. The grey contours in both images are $[2, 3, 4]\,\sigma$~\oeight~emission from ALMA Cycle 2 and Cycle 3 concatenated data \citep{yi23}. The white contours in right panel are $[1, 2, 5, 10, 20, 28]\,\sigma$~\oiii$\lambda4960, 5008$ emission measured from NIRSpec IFU spectroscopy. All the contours are drawn in a per-pixel way and the $1\,\sigma$ noise levels are the standard deviations measured after $3\,\sigma$~clipping in the entire FoVs of the images. The dark grey circles in the bottom right corner of both images indicate the beam size of ALMA Cycle 2+3 observations, which is $0\farcs17\times0\farcs16$. The green circles in both images show PSF FWHMs of F115W and F444W, which are $0\farcs066$~and $0\farcs161$~\citep{Finkelstein}, respectively. The white circle in the right panel shows the PSF FWHM of the NIRSpec spectrum at the wavelength range of optical [O\,{\sc iii}] doublet, which is $0\farcs195$ measured from simulated PSF using STPSF \citep{Perrin14}.}
    \label{spatial distribution}
\end{figure*}

We compare the spatial distribution of UV continuum and \oeight~emission by plotting the \oeight~emission from combined ALMA Cycle 2 and Cycle 3 dataset on F115W image (left panel of Fig. \ref{spatial distribution}). We also compare the spatial distribution of optical and FIR [O\,{\sc iii}] emission by plotting the ALMA \oeight~contours and optical [O\,{\sc iii}] doublet contours derived from NIRSpec IFS on F444W image (right panel of Fig. \ref{spatial distribution}). 

In both panels of Fig. \ref{spatial distribution}, the southern clump of \oeight~emission is spatially aligned with the peak positions of UV continuum and optical [O\,{\sc iii}] emission. In the right panel, the spatial distributions of optical [O\,{\sc iii}] and the emission in F444W are very similar, as the optical [O\,{\sc iii}] emission contributes to the majority of light in F444W. Additionally, the northern clump of \oeight~emission exhibits spatial alignment with the diffuse northern emission of F444W and optical [O\,{\sc iii}].

\citet{yi23} argued that \oeight~emission in this galaxy may have two to three clumps. An isolated northwest clump detected at $\sim3\,\sigma$ may be real or attributed to noise. In this work, we found this isolated clump appears spatially aligned with the $1\,\sigma$ contour of optical [O\,{\sc iii}] emission, implying this northwest clump might be real. Both this clump and the northern clump of \oeight~emission could enfold some obscured or dim star-forming regions that are invisible in F115W. \citet{yi23} found that the northern clump consists of two velocity components, one is blue-shifted and the other one is red-shifted with a velocity separation of $\sim380\,\rm{km}\,\rm{s}^{-1}$. This exceeds the width of narrow \oiii5008 but remains narrower than the broad \oiii5008. Additionally, both the southern and northwest clumps display blue-shifted velocities with FWHMs $\gtrsim250\,\rm{km}\,\rm{s}^{-1}$. These velocity shifts and broadening components may originate from stellar winds or supernova explosions. 

The \oeight~emission shows a clumpy structure but the optical [O\,{\sc iii}] appears smoothly distributed. This is likely due to different critical densities of \oeight~emission and optical [O\,{\sc iii}] emission ($5.1\times10^2\,\rm{cm}^{-3}$ and $6.8\times10^5\,\rm{cm}^{-3}$, respectively; \citealt{osterbrock}) and the inhomogeneous density distribution within the galaxy's H{\,\sc ii} regions. In particular, the valley between the southern and northern clumps of FIR [O\,{\sc iii}] may represent a region where the electron density exceeds the critical density of \oeight~emission, thus suppressing the FIR [O\,{\sc iii}] emission there.  Alternatively, the seemingly clumpy structure of FIR [O\,{\sc iii}] emission may result from observational effects. ALMA's high angular resolution is possible to make smooth disks look clumpy \citep{gullberg18}, suggesting that the clumpy morphology of FIR [O\,{\sc iii}] may be artefact. Furthermore, since the peak S/N of FIR [O\,{\sc iii}] is only $4.8\,\sigma$, much smaller than that of optical [O\,{\sc iii}], the undetected diffuse emission in the outer region of FIR [O\,{\sc iii}] could be attributed to the insufficient sensitivity. Deeper ALMA observations are of importance to probe the diffuse emission of FIR [O\,{\sc iii}]. 

The UV continuum exhibits an elongated morphology along the east-west direction, suggesting that we are likely observing the galaxy disk from edge-on direction, and the more extended and diffuse continuum in longer wavelengths as well as the \oiii$\lambda$4960, 5008 emission form an ellipsoidal or spherical ionized halo surrounding the disk. However, this structure could also be caused by decreasing resolutions at longer wavelengths. The UV continuum in F115W and F150W looks clumpy too (Fig. \ref{image}); thus, it is also possible that our target is a chain galaxy \citep{elmegreen04}. Both the UV continuum and \oeight~emission show a tail-like structure extending toward the west (Fig. \ref{image}; Fig. \ref{spatial distribution}), likely indicating the presence of diffuse star-forming regions adjacent to the central star-forming clusters. The elongated, clumpy and tail-like structures of the emission may also result from galaxy-merging events, which boost the current bursty star formation. Indeed, at the redshift of SXDF (i.e. $z\sim7$), recent JWST observations have shown a high prevalence of clumpy galaxies and/or close merging companions (e.g., \citealt{harikane25}).

\subsection{Potential Old stellar Population}
\label{underlying population}
The mass-weighted mean ages from all SFH models are $\sim1-2$ Myr (Table \ref{SED results}), which is even smaller than the lifetime of the most massive stars \citep{buzzoni02}. However, the observed emission lines and the SED fitting suggest a metallicity of $\sim0.1-0.2\, \rm{Z}_\odot$. Even an inferred age of $\sim10$ Myr since onset of SF remains challenging to reconcile with the derived chemical abundance, suggesting potential presence of old stellar populations responsible for the metal enrichment in this galaxy.

\subsubsection{Reconstruct Past Star Formation History with SED Fitting}
We first attempt to reconstruct old stellar assembly histories using SED fitting. As mentioned in \S\ref{parameter setting}, when fitting with the non-parametric SFH with a continuity prior, the choice of the earliest age bin may influence both the onset time of SF and the output of stellar mass. Adopting an earliest age bin at 20 Myr yields age and stellar mass consistent with results from parametric models. In contrast, setting the earliest age bin at $z=20$ results in a mass-weighted mean age of $\sim10$ Myr and a timescale since onset of SF of $10^{+290}_{-7}$ Myr. The large upper uncertainty implies the possible presence of previous SF activities (bottom left panel of Fig. \ref{sfh}).

Furthermore, fitting a non-parametric model with a Dirichlet prior and a concentration parameter of $\alpha < 1$ allows for sharp changes in SFR, such as bursts, rapid quenching and rejuvenation events, while $\alpha \geq 1$ suppresses abrupt SFR variations, similar to the effect of a continuity prior \citep{leja19}. Thus, to explore if drastic SF activities were present in the past SFH, we performed SED fitting employing a non-parametric model with a Dirichlet prior and $\alpha = 0.2$ using Dense Basis \citep{iyer19}. 

We only input NIRCam photometries in this fitting. Since old populations primarily contribute to fluxes at rest-frame $\lambda\gtrsim4000\,$\AA, excluding constraints from [O\,{\sc iii}] emission lines and undetected dust continuum has minimal impact on SFH reconstruction. We fixed metallicity, redshift and dust attenuation by adopting very tight priors centered on values derived from Bagpipes fittings, allowing the maximum lookback times to be $N=3$.

As a result, the derived stellar mass of $\log(M_\ast/\rm{M}_\odot) = 8.59^{+0.68}_{-0.25}$ is consistent with the Bagpipes results, and the instantaneous SFR of $\log(\rm{SFR}/M_\odot\,yr^{-1})=1.62^{+0.17}_{-0.18}$ also agrees well with the SFR inferred from H$\beta$ emission. The reconstructed SFH suggests a flat, moderate star formation beginning from soon after the Big Bang to the recent burst (Fig. \ref{iyer_sfh}), which may imply the presence of old populations. Even though we defined a low $\alpha$ of 0.2, the median distribution of SFH implies no sharp SFR changes in the past.

\subsubsection{Maximum Old Stellar Mass}
We further try to estimate the maximum possible mass of the potential old stellar populations. Given that the flux densities of stellar populations with ages $\gtrsim 0.4$ Gyr peak at rest-frame $\sim4000\,$\AA \citep{ssp}, we utilized the discrepancy between observed and modeled flux in F356W from the $\tau$ SFH model fit to estimate the maximum possible mass of potential old stellar populations. We assumed that this flux discrepancy is completely due to the contribution of the old populations. To measure their mass, we performed SED fitting while assuming a bursty star formation occurred at $z=20$, followed by a passive evolution, because the earliest galaxy formation is expected to start around this time \citep{robertson22}. We fitted a constant SFH model, where the onset of star formation is at the cosmic age of $0.175\,\rm{Gyr}$ (corresponding to $z=20$) and the duration is $10\,\rm{Myr}$. We assumed a 10 per cent error floor for the flux of old population in F356W. We also assumed the metallicity of the gas cloud at $z\sim20$ is as low as the most metal-poor galaxy so far found in the local Universe ($\sim1/50\,\rm{Z}_\odot$, I Zw 18, \citealt{Izotov1999}) and the dust attenuation at $z\sim20$ is 0.01 mag. Consequently, the estimated maximum underlying old stellar mass is $\log(M_{\ast}^{\rm{old}}/\rm{M}_\odot) = 9.50\pm0.04$, which we adopt as an upper limit of $\log(M_{\ast}^{\rm{old}}/\rm{M}_\odot) < 9.5$. This constraint is 1.2 dex more stringent than that of \citet{inoue16}, who derived the potential old mass using flux upper limit in Spitzer IRAC $3.6\,\micron$, much brighter than F356W. Additionally, assuming a bursty SF occurred at a later time ($z = 10$) with a higher metallicity ($Z/\rm{Z}_\odot = 0.1$) reduces the potential old mass by only $\sim0.2$ dex. After incorporating the flux contribution from old populations at the effective wavelength of F356W, the discrepancy between the modeled and observed flux in F356W is resolved. The presence of the old populations can be supported by FirstLight simulations, which predicted that the typical stellar mass of $z\sim6$ starburst galaxies above the main sequence with $M_{\rm{UV}}\sim-22$ is approximately $\log(M_\ast/\rm{M}_\odot)\sim9.3-9.5$ \citep{Ceverino19}. Meanwhile, \citet{langan20} and \citet{nakazato23} reported that simulated starbursts at $z = 7$ with metallicities comparable to SXDF could have stellar masses up to $\sim10^9\,\rm{M}_\odot$.
\subsubsection{Use MIRI to Trace Old Populations}
\label{miri}
The presence of the potential old stellar populations can be confirmed with future MIRI observations. For our target, MIRI F770W is the most useful band to trace old populations, because H$\alpha$ emission may outshine the stellar continuum in F560W, and F770W offers the second-best sensitivity while minimizing contamination from emission lines.

To predict the observed flux in F770W, we combined the posterior spectrum of young populations derived from SED fitting with the model spectrum of potential old populations, integrating over the F770W bandpass. This yielded an expected flux density of $0.23\pm0.02\,\mu\rm{Jy}$. Given that the $10\,\sigma$ sensitivity limit of F770W is $0.24\,\mu\rm{Jy}$ for a 10 ksec exposure \footnote{JWST User Documentation: \url{https://jwst-docs.stsci.edu}}, the predicted S/N with a two-hour integration would reach $\sim8.0\,\sigma$. It would be enough to securely confirm the existence of old populations, and to constrain further the SFH of this source up to very early epochs ($z\sim20$). Conversely, in the absence of old populations, the expected flux density decreases to $0.17\pm0.02\,\mu\rm{Jy}$, corresponding to a predicted S/N of $\sim5.9\,\sigma$ for the same exposure time. In summary, the brightness of emission detected in F770W of future MIRI observations can confirm the presence of old stellar populations, and further constrain the stellar mass and SFH of our target.

Among MIRI imaging filters, F1000W and F1280W have the third- and fourth-best sensitivities, respectively. Appying the same calculation approach described above, the predicted S/Ns are only $3.7\,\sigma$ (F1000W) and $\sim1.1\,\sigma$ (F1280W) for a two-hour integration, assuming the presence of old populations. Thus, F770W is the most viable band to investigate the potential old stellar populations in our target. Although the predicted S/N is modest in F1000W and very low in F1280W, these observations may still be valuable for constraining the old stellar populations and tracing hot dust.

Recent JWST studies have reported young stellar ages (a few to dozens of Myr) in high-$z$ galaxies exhibiting bursty star formations (e.g. \citealt{Endsley23faintuv,tang23,chen23,looser23, sugahara25}). Such short timescales appear insufficient to account for the observed chemical abundance, implying the presence of underlying old stellar populations in these galaxies. 
JWST NIRCam observations of a local extremely metal-poor galaxy I Zw 18, which is considered as a local analogue of high-$z$ galaxies, have identified evidence of stellar populations older than 1 Gyr in this system \citep{Bortolini24}. This emphasizes the importance of rest-frame NIR photometry on probing old stellar populations.
Moreover, MIRI imaging is crucial for studying cosmic stellar mass density at high-$z$ \citep{papovich23}. Utilizing MIRI's capability, we can probe the stellar populations at much earlier cosmic epochs than those corresponding to the observed redshifts of high-$z$ galaxies.

\subsection{Ionization and ISM Condition}
\label{ionization and ism condition}
Since F115W's bandpass covers rest-frame $1500\,$\AA, we measured the absolute UV magnitude using the observed total flux density in this filter. Following the dust extinction correction approach outlined in \S\ref{strong line method}, we obtained $M_{\rm{UV}} = -22.1\pm0.2$, consistent with previous measurement in \citet{shibuya12}. Moreover, we derived the rest-frame EW of [O\,{\sc iii}]+H$\beta$ using the model continuum level obtained from the posterior distribution of SED fitting, yielding an extremely high value of $3700\pm370\,$\AA. It places our target at the upper end of JWST-detected $z\sim6.5-8$ and $z\sim6-9$ galaxies from CEERS and JADES surveys, respectively, whereas it is comparable with sources exhibiting Balmer jumps in their modeled SEDs (\citealt{Endsley23faintuv, endsley_jades}). We also computed the EW of Ly$\alpha$ emission. We estimated the continuum level of Ly$\alpha$ using the UV slope derived from modeled SED (Table \ref{many result}) and the observed F115W flux density. Given the Ly$\alpha$ flux of $1.9^{+2.5}_{-0.9}\times10^{-17}\,\rm{erg}\,\rm{s}^{-1}\rm{cm}^{-2}$ \citep{shibuya12}, we derived an EW of $30^{+39}_{-14}\,$\AA, consistent with the value obtained using continuum level from UKIRT $J$ band observations \citep{yi23}.

We also estimated the ionizing photon production rate employing equation (4) from \citet{reddy23highz}, which assumes no dust absorption of ionizing photons or leakage into the intergalactic medium (IGM). By substituting the predicted intrinsic H$\alpha$ luminosity into the equation, we obtained $\log(Q/\rm{s}^{-1}) = 54.9\pm0.1$. We further determined ionizing photon production efficiency following $\xi_{\rm{ion}} = Q / L_\nu$, where $L_\nu$ represents the intrinsic rest-frame UV luminosity density at $1500\,$\AA. We determined $L_\nu$ using observed flux density in F115W after applying corrections for both cosmological expansion and dust extinction. We corrected for dust extinction following the same procedure outlined in \S\ref{strong line method}. The obtained value is $\log(\xi_{\rm{ion}}/\rm{erg}^{-1}\,Hz) = 25.44\pm0.15\,$, which agrees well with the measurement from \citet{inoue16}. The $\xi_{\rm{ion}}$ is also consistent with galaxies at $z\gtrsim6$ identified by JWST spectroscopic observations \citep{Javer_jd1, Javier_gnz11} and strong LAEs at $5<z<8$ observed by JWST NIRSpec \citep{Heintz25}.

The ionization parameter derived from SED fitting ($\log U \sim -1.6$) is higher than those of local H\,{\sc ii} regions \citep{kewley02}, and consistent with the highest estimated values of $z\sim1.3-2.3$ SFGs within uncertainties \citep{masters14}. The marginally detected \neiii3870 and [O\,{\sc ii}] doublet emission enable the estimation of $\log U$ through O32 (\oiii$\lambda$4960, 5008/\oii$\lambda$3727, 3730), Ne3O2 (\neiii3870/\oii$\lambda$3727, 3730) and R23 ((\oiii$\lambda$4960, 5008 + \oii$\lambda$3727, 3730)/H$\beta$) proxies. The derived dust-corrected line ratios are $\rm{O32} = 18.3^{+6.7}_{-4.8}$, $\log\rm{Ne3O2} = 0.06\pm0.20$ and $\rm{R23} = 15.1^{+5.0}_{-3.7}$ \footnote{\citet{cameron23} adopted O32 = log (\oiii5008/\oii$\lambda$3727, 3730), which yields a value of $\log\rm{O}32 = 1.17\pm0.14$.}. The measured O32, Ne3O2 ratios and the $\log U$ derived from SED fitting place our target at the upper ranges of $z\sim2.7-6.3$ and $z\sim7-9$ SFGs selected from JWST CEERS survey. Nevertheless, our derived $\log U$ aligns with the model-inferred $\log U$ of $z\sim2.7-6.3$ samples with comparable O32 and Ne3O2 ratios, and the $\log U$ vs. $\rm{O32}$, $\rm{O32}$ vs. $\rm{Ne3O2}$ and $\rm{O32}$ vs. EW([O\,{\sc iii}]+H$\beta$) of our target show consistency with $z>7$ LAEs (\citealt{reddy23highz, tang23}). Moreover, while the R23 ratio places our target at the upper range of $z\sim5.5-9.5$ JWST JADES sources, both O32 and Ne3O2 ratios match their stacked values \citep{cameron23}. 

Recent studies by \citet{reddy23highz} and \citet{reddy23lowz} have established significant linear correlations between $\log U$ and SFR surface density ($\Sigma_{\rm{SFR}}$), as well as between O32 and $\Sigma_{\rm{SFR}}$, for star-forming galaxies at $z\sim2.7-6.3$ and $z\sim1.6-2.6$. We measured $\Sigma_{\rm{SFR}}$ of our target using equation (5) from \citet{reddy23highz}:
\begin{equation}
    \Sigma_{\rm{SFR}} = \frac{\rm{SFR}}{2\pi{\it{R}}^2_{\rm{eff}}}.
\end{equation}
We derived two estimates: (1) $\Sigma_{\rm{SFR}} = 32.6\pm17.2\,\rm{M}_\odot\,\rm{yr}^{-1}\,\rm{kpc}^{-2}$ using instantaneous SFR derived from SED fitting, and (2) $\Sigma_{\rm{SFR}} = 8.2\pm4.8\,\rm{M}_\odot\,\rm{yr}^{-1}\,\rm{kpc}^{-2}$ using H$\beta$-based SFR, while adopting the effective radius of narrow component of \oiii5008 emission (Table \ref{many result}). Both measurements place our target above the established correlations of O32 vs. $\Sigma_{\rm{SFR}}$ and $\log U$ vs. $\Sigma_{\rm{SFR}}$, whereas they align with the most extreme $2.7<z<6.3$ systems showing similarly elevated O32 ($\log U$) and $\Sigma_{\rm{SFR}}$ (see Fig. 9 and Fig. 10 in \citealt{reddy23highz}). More samples at $z\gtrsim7$ with both O32 ($\log U$) and $\Sigma_{\rm{SFR}}$ measurements are necessary for studying their correlations in early galaxies. Additionally, our target's metallicity and $\log U$ show better agreement with $z\gtrsim7$ samples than with lower redshift samples (see Fig. 7 in \citealt{reddy23highz}).

\subsection{Age-Mass and Mass-Metallicity Relations of ALMA-detected and JWST-selected Galaxies}
\label{evolutionary pathway}
Prior to the JWST-era, star-forming galaxy candidates at $z>6$ selected by past telescopes and successfully detected by ALMA follow-up observations are generally bright in rest-frame UV or benefited from strong gravitational magnification \citep[e.g.,][]{inoue16, hashimoto18, Tamura19, bouwens22}, given the shallower detection limits of previous instruments and competitive ALMA time. In this section, we investigate early galaxy evolution by comparing the stellar mass, age, $M_{\rm{UV}}$ and metallicity of galaxies detected by both ALMA and JWST (hereafter ALMA-detected galaxies) and galaxies selected by JWST observations. We compiled a sample of 12 ALMA-detected galaxies at $z>6$ from the literature (including our target). We compared these galaxies with JWST-selected galaxies from JADES and CEERS programs in terms of stellar mass, age and $M_{\rm{UV}}$ (Fig. \ref{age_mass}). We further included 12 galaxies from REBELS program, other four ALMA-detected galaxies, and galaxies from JWST ERO and GLASS programs to compare their mass-metallicity relations (Fig. \ref{mass_Z}). More details of the plotted data are presented in Appendix \ref{alma_detected_appendix}.

\subsubsection{Age-Mass Relation}
\begin{figure*}
    \includegraphics[width = 1.0\textwidth]{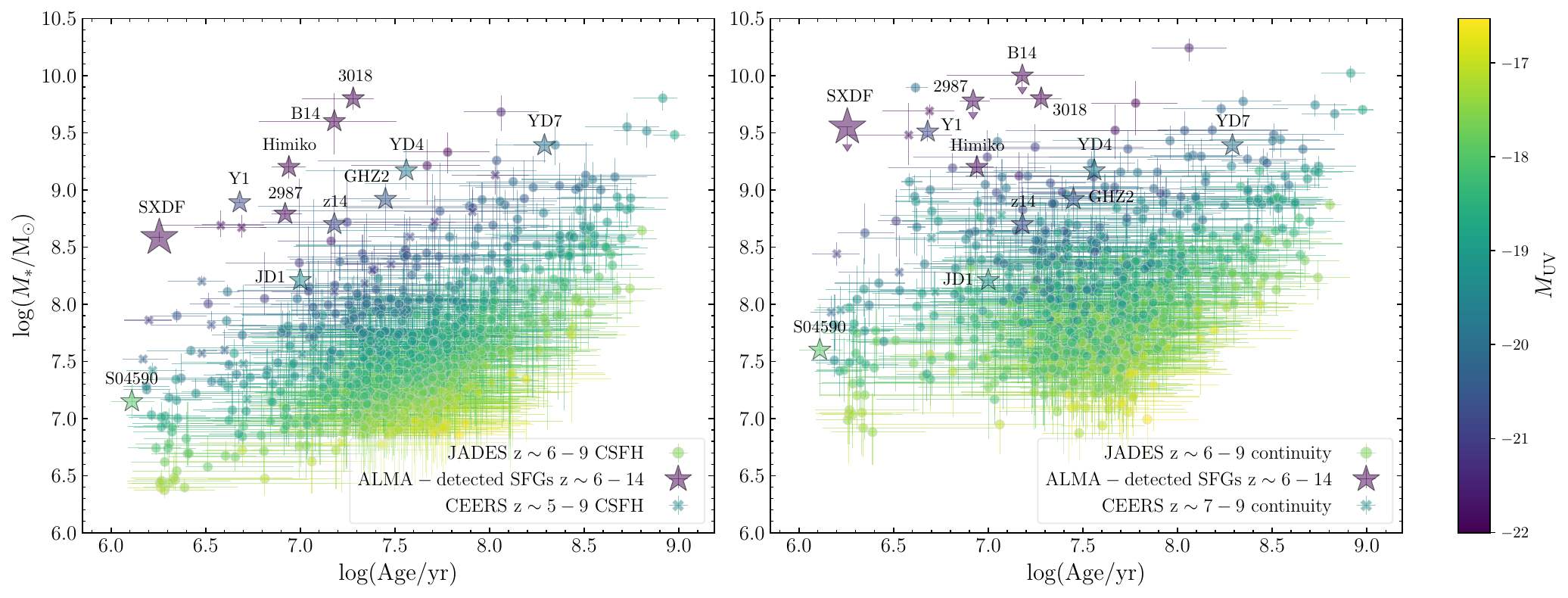}
    \caption{Comparison of ALMA-detected star-forming galaxies and JWST-selected galaxies at $z>6$. The circles indicate galaxies selected by JADES program \citep{endsley_jades}, the crosses represent galaxies from CEERS program \citep{tang23, chen23}, and the stars show SFGs detected by both ALMA and JWST. Specifically, these galaxies are: our target (highlighted by the largest star); B14-65666 \citep{sugahara25}; COS-2987030247 \citep{mawatari25}; COS-3018555981 \citep{harikane25}; Himiko \citep{harikane25}; MACS0416-Y1  \citep{ma24, Harshan24}; MACS1149-JD1 \citep{stiavelli23}; A2744-YD4 and -YD7 \citep{hashimoto23, morishita23}; JADES-GS-z14-0 \citep{carniani24_2, schouws25, Helton25}; S04590 \citep{Heintz_z8, fujimoto24_z8.5} and GHZ2/GLASS-z12 \citep{zavalaFIRO3, zavalaMIRI}. For JWST galaxies, the horizontal axes indicate stellar age derived by SED fitting with a constant SFH (CSFH), while the vertical axes show stellar mass inferred with either a CSFH (left panel) or a non-parametric continuity SFH (right panel).
    More detailed information of plotted data is presented in Appendix \ref{alma_detected_appendix}. The data points are color-coded by $M_{\rm{UV}}$.
    }
    \label{age_mass}
\end{figure*}

In Fig. \ref{age_mass}, most of the ALMA-detected galaxies occupy the upper left region of the panels, exhibiting higher stellar mass, younger age and brighter UV magnitude, compared to JWST-selected galaxies. Especially, when compared to the constant SFH (CSFH)-derived mass of JWST-selected sources (left panel), the ALMA-detected galaxies have significantly larger mass at similar ages. Although the non-parametric continuity SFH may yield more mass particularly for young systems, the ALMA-detected galaxies remain more massive than the majority of the JWST-selected galaxies (right panel). Particularly, the age of our target is situated at the lowest region compared to all the other galaxies, but it is significantly more massive than other youngest galaxies.

There are three outliers, S04590, MACS1149-JD1 and A2744-YD7. The mass of S04590 is very low ($\sim10^{7.2}\,\rm{M}_\odot$), and the mass of MACS1149-JD1 ($\sim10^{8.2}\,\rm{M}_\odot$) is consistent with JWST-selected galaxies at similar ages (right panel of Fig. \ref{age_mass}). Both systems benefit from much stronger gravitational lensing effect than other magnified sources, which enhanced their chance of being detected by ALMA. Additionally, the age of YD7 ($\sim195$ Myr) is the oldest among ALMA-detected samples and close to the upper age limit of JWST-selected sources. This is because YD7 has two components, one has young bursting populations (YD7-W), while the other one has old star-forming activities and is recently quenched \citep[YD7-E;][]{Witten25}. Thus, these three galaxies are not contradictory with the overall trend showing ALMA-detected galaxies are typically more massive and younger than JWST-selected galaxies.

We have to note, however, that this trend looks very significant because most of the JWST-selected samples in Fig. \ref{age_mass} are from JADES program, which has a median $M_{\rm{UV}}$ of $-18.3$ for galaxies detected by \citet{endsley_jades}. This is much fainter than most ALMA-detected galaxies and fainter than some other JWST projects like CEERS (median $M_{\rm{UV}} = -19.8$ for samples in \citealt{nakajima23}), considering the positive correlation between $M_{\rm{UV}}$ and stellar mass \citep[eg.,][]{endsley_jades}. In Fig. \ref{mass_Z}, with the complementary data from other programs, the mass of ALMA-detected and JWST-selected galaxies becomes more consistent; however, the mass of most ALMA samples still occupies the higher value end.

The young but more massive nature of ALMA-detected galaxies at $z>6$ implies that these galaxies may have undergone more efficient mass assembly  processes in their formation and evolution histories. We calculated the clumpy and merger fractions of the ALMA samples. B14-65666, COS-2987, COS-3018, Himiko, CR7, MACS0416-Y1, MACS1149-JD1, BDF-3299 and COSMOS24108 show clumpy structures in UV continuum and evidence of past or ongoing merger activities \citep{sugahara25, mawatari25, scholtz24, harikane25, kiyota25, Harshan24, Marconcini_jd1, venturi24}. Despite not being clumpy in the stellar continuum, A2744-YD4 also shows evidence of mergers, hinted by a flat metallicity gradient \citep{venturi24}. Furthermore, A2744-YD1, -YD4 and -YD7 are in a protocluster core where strong environmental effects can accelerate galaxy evolution \citep{hashimoto23}, and they are predicted to be at the verge of galaxy mergers by cosmological simulations \citep{nakazato24}. The clumpy structure of our target shown in F115W may also result from merging events. The extended, blueshifted [C\,{\sc ii}] emission in S04590 could be explained by merger activities as well \citep{Heintz_z8}. Overall, the clumpy and merger (with strong evidence) fractions of our compiled 16 ALMA-detected SFGs are 69 percent and 63 percent, respectively. The clumpy fraction is consistent with bright ($M_{\rm{UV}}<-21.5$) galaxies at $z\sim7$ compiled by \citet{harikane25} ($66\pm14$ percent), as our galaxies are also characterized by bright UV color and massive stellar mass. The merger fraction of our ALMA samples is much larger than the major merger fraction determined by JWST-selected galaxies at $z>6$ \citep[$\sim0.2$;][]{duan24}, suggesting that galaxy interactions may play an important role in accelerating the mass assembly in bright and massive galaxies like ALMA-detected ones at $z>6$.

Furthermore, while the JWST-selected faint, less massive and young galaxies may have formed recently prior to the observed redshifts, the more massive ALMA-detected galaxies with similarly young ages are more likely to have experienced past star-forming activities that began dozens or even hundreds of Myr ago, as shown by the analysis of potential old stellar populations in our target. The ALMA-detected galaxies in the reionization epoch, which represent a young but massive and bright galaxy population, likely underwent more complex evolutionary pathways, such as the more frequent merging activities discussed above.

\subsubsection{Mass-Metallicity Relation}
\begin{figure*}
    \includegraphics[width = 0.8\textwidth]{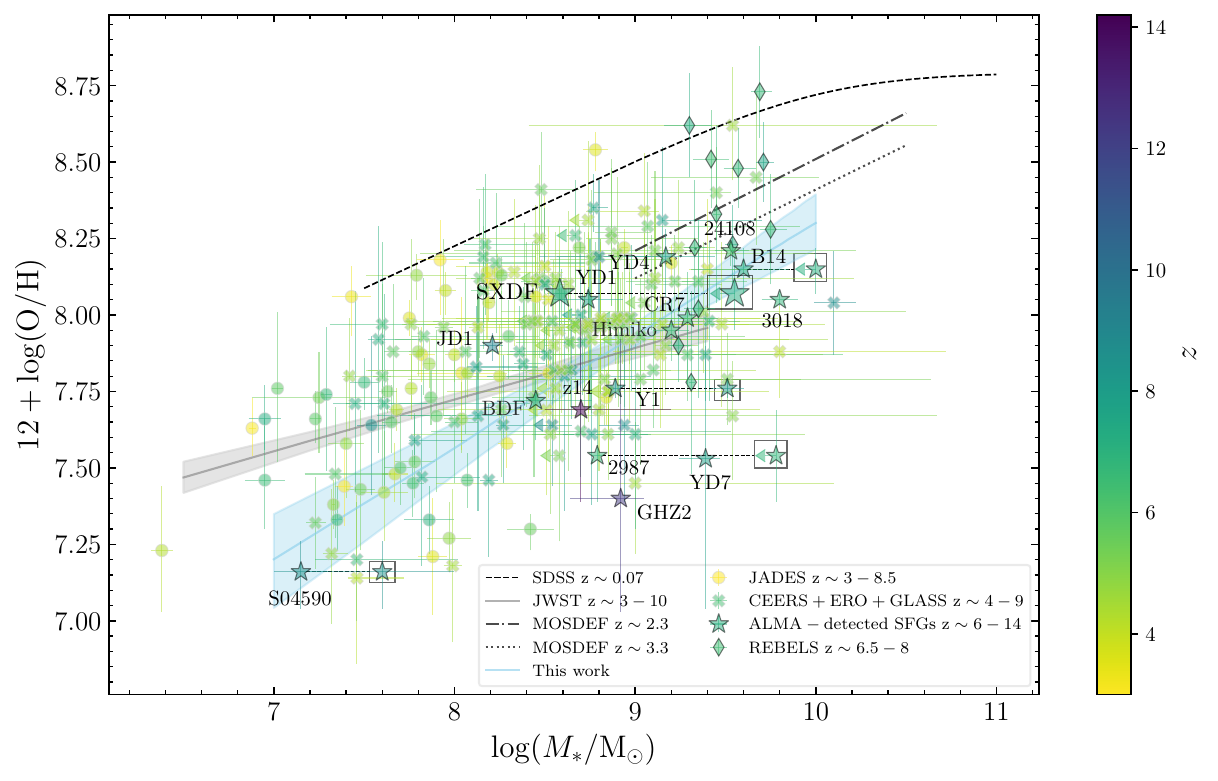}
    \caption{Mass-metallicity relation of JWST- and ALMA-detected galaxies. Our target is highlighted by the largest star. The JADES galaxies are from \citet{curti24} and CEERS+ERO+GLASS galaxies are from \citet{nakajima23}. Apart from the ALMA samples plotted in Fig. \ref{age_mass}, we further included four galaxies detected by ALMA at $z>6$: CR7, A2744-YD1, BDF-3299 and COSMOS24108, and 12 REBELS galaxies from \citet{Rowland25}. The stars connected by a black dashed line represent young and total mass (enclosed in a black rectangle). More detailed information of plotted data is presented in Appendix \ref{alma_detected_appendix}. The mass-metallicity relations at $z\sim0, z\sim2-3$ and $z\sim3-10$ are from \citet{curti20, sanders21, curti24}, respectively. The grey shaded region indicates the $1\,\sigma$ uncertainty of the $z\sim3-10$ relation \citep{curti24}. The mass-metallicity relation of $z>6$ ALMA-detected galaxies (including REBELS galaxies) is shown by the blue solid line, and the blue shaded region shows its $1\,\sigma$ uncertainty. The data points are color-coded by redshift.}
    \label{mass_Z}
\end{figure*}

In Fig. \ref{mass_Z}, most of the ALMA-detected galaxies (including our compiled 16 galaxies and REBELS galaxies) show deviations from local samples, consistent with other high-$z$ samples. We conducted a linear orthogonal distance regression fit to ALMA-detected galaxies using the following functional form:
\begin{equation}
    12+\log(\rm{O/H}) = \beta\log(\it{M}_\ast/\rm{10^9\,M}_\odot) + \it{Z}_{\rm{9}},
\end{equation}
where $\beta$ is the slope and $Z_9$ is the metallicity at the normalized stellar mass of $10^9\,\rm{M}_\odot$. We found $\beta = 0.37\pm0.08$ and $Z_9 = 7.94\pm0.04$. The slope agrees well with that reported by \citet{Rowland25} who fitted REBELS data as well. It is steeper than the $z\sim3-10$ and $z\sim4-10$ relations of JWST-selected galaxies reported by \citet{curti24} and \citet{nakajima23}, respectively. Its median value is also larger than the $z\sim2-3$ relations in \citet{sanders21}, the $z\sim3-9.5$ relation in \citet{morishita24} and the $z\sim7-10$ relation in \citet{Heintz23NA}. However, the slope of our samples is consistent with these relations if considering large uncertainties. In fact, our measured slope would be even steeper if considering the potential old stellar mass in some targets or excluding low-mass samples (i.e. S04590 and JD1). This steep slope may be attributed to the fact that the ALMA-detected galaxies at $z>6$ are more massive, but are dominated by young, massive and metal-poor stars.


\section{conclusion}
In this work, we analyzed JWST NIRCam and NIRSpec observations of a galaxy at $z=7.212$, as part of the RIOJA project. We summarize our conclusions as follows.

$1.$ We observed our target using six NIRCam wideband filters at $1-5\,\micron$. The UV continuum in SW filters exhibits an elongated, clumpy morphology with a tail-like structure, suggesting an edge-on disk or a chain galaxy. In contrast, the optical emission is more extended and smoothly distributed, which may result from decreasing resolutions at longer wavelengths, or indicating the presence of optical halo. The \oeight~emission is also clumpy with a tail-like structure extending toward the west. Thus, our target may has experienced galaxy merger events in the past.

$2.$ We conducted NIRSpec IFS observations and identified \oii$\lambda3727, 3730$, \neiii3870, H$\beta$ and \oiii$\lambda$4960, 5008 emission lines at galactic scales. We also detected blended He\,{\sc i} + [Ne\,{\sc iii}] + H$\epsilon$ lines at $\sim3970\,$\rm{\AA} and H$\gamma$ in a smaller aperture. We detected a broad component of \oiii5008 with an extent of $r_{\rm{e}} = 1.86\pm0.63\,$kpc, indicating the presence of prominent ionized gas outflows. The total mass loading factor may be larger than one. However, due to the large uncertainties of the estimates, we can not conclude whether the outflow-driven mechanisms can quench the star formation in our target effectively.

$3.$ We performed SED fitting using NIRCam photometries, line fluxes of \oiii$\lambda4960, 5008$ and \oeight~emission lines, as well as dust continuum non-detections obtained from ALMA. Our results show that our target is dominated by young stellar populations and undergoing a bursty SF phase. The modeled SED exhibits a prominent Balmer jump, indicating the presence of strong nebular continuum. The derived ionization parameter, O32 and Ne3O2 line ratios, and EW([O\,{\sc iii}]+H$\beta$) are higher than the typical values of high-$z$ SFGs, but consistent with those of $z>7$ LAEs. 

$4.$ We found the inferred young stellar ages from SED fitting are insufficient to account for the observed chemical abundance of $\sim0.2\,\rm{Z}_\odot$, implying the presence of underlying old populations. The estimated maximum old stellar mass is $\log(M_{\ast}^{\rm{old}}/\rm{M}_\odot)<9.5$. MIRI observations are essential for investigating the potential old stellar populations and constraining the stellar mass and SFH of high-$z$ SFGs exhibiting young stellar ages. MIRI F770W is the most viable band to probe the old populations for our target.

$5.$ We compared the spatial distribution of \oeight~emission, \oiii$\lambda$4960, 5008 emission and UV continuum. The southern clump of \oeight~emission is spatially aligned with the peak positions of UV continuum and optical [O\,{\sc iii}] emission, while the northern clump of \oeight~emission is aligned with the diffuse northern emission of optical [O\,{\sc iii}]. Different structure of optical and FIR [O\,{\sc iii}] could be attributable to different critical densities of these lines and inhomogeneous density distribution in the galaxy. The clumpy structure of FIR [O\,{\sc iii}] may also be artefact resulted from the high-angular resolution observation of ALMA. Deeper ALMA observations would be necessary to trace the diffuse emission of FIR [O\,{\sc iii}].

$6.$ We derived dynamical mass, molecular gas mass and H\,{\sc i} gas mass using [C\,{\sc ii}] observations obtained from ALMA. We estimated gas depletion time using the derived gas mass and SFRs, yielding a $t_{\rm{depl}}$ of $\sim114$ Myr or $\sim445$ Myr, under different SFR assumptions. This implies that our galaxy may be quenched at $z\sim6.5$ or $z\sim5$, suggesting that our target may be one of the progenitors of observed massive quiescent galaxies at $z\sim4-5$.

$7.$ We compared the age-mass and mass-metallicity relations of JWST-selected galaxies and star-forming galaxies detected by both JWST and ALMA at high redshifts. We found that, the ALMA-detected galaxies at $z>6$ are relatively more massive, younger, and their mass-metallicity relation may have a steeper slope. Therefore, the ALMA-detected galaxies at $z>6$, which represent a young but massive and UV-bright galaxy population, may have undergone more efficient mass assembly processes in their formation and evolution histories, as suggested by their high clumpy and merger fractions (69 percent and 63 percent, respectively). The presence of potential old stellar populations (e.g., in our target) also suggests that star formation in the ALMA sources may have begun at a much earlier time, compared to young but less massive galaxies.

\section*{Acknowledgements}
We thank the anonymous referee for precious comments that have greatly improved the quality of this paper.
Y.W.R. was supported by JSPS KAKENHI Grant Number 23KJ2052. A.K.I. and K.M. was supported by JSPS KAKENHI Grant Numbers 23H00131 and 24H00002.
T.H. and K.M. was supported by Leading Initiative for Excellent Young Researchers, MEXT, Japan (HJH02007) and by JSPS KAKENHI Grant Number 22H01258. 
A.K.I. and Y.S. was supported by NAOJ ALMA Scientific Research Grant number 2020-16B. 
Y.T. was supported by JSPS KAKENHI Grant Number 22H04939.
J.A.M., A.C.G., L.C. and S.A. acknowledge support by grant PIB2021-127718NB-100 from the Spanish Ministry of Science and Innovation/State Agency of Research (MCIN/AEI/10.13039/501100011033) and by “ERDF A way of making Europe”. Y.N. was supported by JSPS KAKENHI Grant Number 23KJ0728.
D.C. is supported by the Ministerio de Ciencia, Innovación y Universidades under research grants PID2021-122603NB-C21 and CNS2024-154550.
K.M. was supported by JSPS KAKENHI grant No. 20K14516.
The project that gave rise to these results received the support of a fellowship from the “la Caixa” Foundation (ID 100010434). The fellowship code is LCF/BQ/PR24/12050015.
L.C. acknowledges support from grants PID2022-139567NB-I00 and PIB2021-127718NB-I00 funded by the Spanish Ministry of Science and Innovation/State Agency of Research  MCIN/AEI/10.13039/501100011033 and by “ERDF A way of making Europe”. C.B.P. acknowledges the support of the Consejería de Educación, Ciencia y Universidades de la Comunidad de Madrid through grants No. PEJ-2021-AI/TIC-21517 and PIPF-2023/TEC29505. J.A.M. and C.B.P. acknowledge support by grant PID2024-158856NA-I00 from the Spanish Ministry of Science and Innovation/State Agency of Research MCIN/AEI/10.13039/501100011033 and by “ERDF A way of making Europe”.

This paper makes use of the following ALMA data: ADS/
JAO.ALMA\#2013.1.01010.S, JAO.ALMA\#2015.A.00018.S, JAO.ALMA\#2012.1.00374.S, JAO.ALMA\#2013.A.00021.S, JAO.ALMA\#2019.1.01634.L and JAO.ALMA\#2021.1.01323.S. ALMA is a partnership of ESO (representing its member states), NSF (USA) and NINS (Japan), together with NRC (Canada), MOST and ASIAA (Taiwan), and KASI (Republic of Korea), in cooperation with the Republic of Chile. The Joint ALMA Observatory is operated by ESO, AUI/NRAO and NAOJ.
This work has made use of data from the European Space Agency (ESA) mission
{\it Gaia} (\url{https://www.cosmos.esa.int/gaia}), processed by the {\it Gaia}
Data Processing and Analysis Consortium (DPAC,
\url{https://www.cosmos.esa.int/web/gaia/dpac/consortium}). Funding for the DPAC
has been provided by national institutions, in particular the institutions
participating in the {\it Gaia} Multilateral Agreement.
This research has made use of NASA’s Astrophysics Data System.
This research made use of Photutils, an Astropy package for detection and photometry of astronomical sources (\citealt{bradley22, bradley24}).
$\it{Software}$: Astropy (\citealt{astropy1, astropy2, astropy3}), CASA \citep{casa}, NumPy \citep{numpy}, Matplotlib \citep{matplotlib}, SciPy \citep{scipy}, Bagpipes \citep{bag18}, Cloudy \citep{cloudy}, STPSF \citep{Perrin14}, Dense Basis \citep{iyer19}.
\section*{Data Availability}
The data that support the findings of this study are available from the corresponding author, upon reasonable request.
\bibliographystyle{mnras}
\bibliography{draft}
\appendix
\section{SED Fitting Results With Varying SFH Models}
\begin{figure*}
    \includegraphics[width=\textwidth]{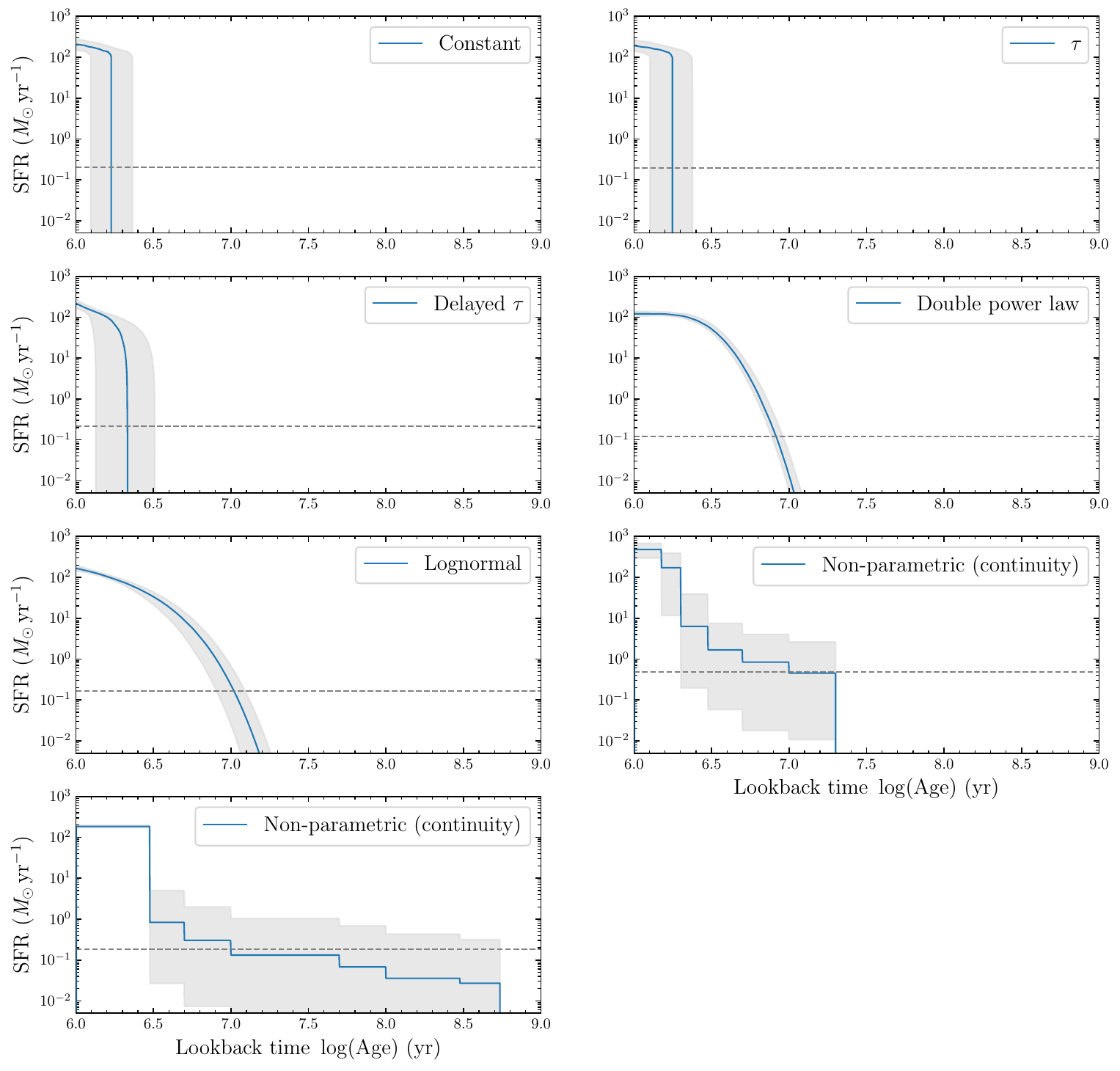}
    \caption{SFHs derived from the posterior output samples of SED fittings using BAGPIPES with different assumptions of SFH models. The non-parametric model from \citet{leja19} assembled in BAGPIPES has a continuity prior. The bottom right panel demonstrates reconstructed SFH where the earliest age bin is set at 20 Myr, while the bottom left panel assumes a starting point at $z = 20$. The blue lines and gray shaded regions represent the 50th and 16th-84th percentiles of posterior output samples. We define onset of SF as the point when SFR exceeds $0.1$ per cent of the peak SFR. The grey dashed lines indicate $0.1$ per cent of the median peak SFR.}
    \label{sfh}
\end{figure*}

\begin{figure}
    \includegraphics[width=0.47\textwidth]{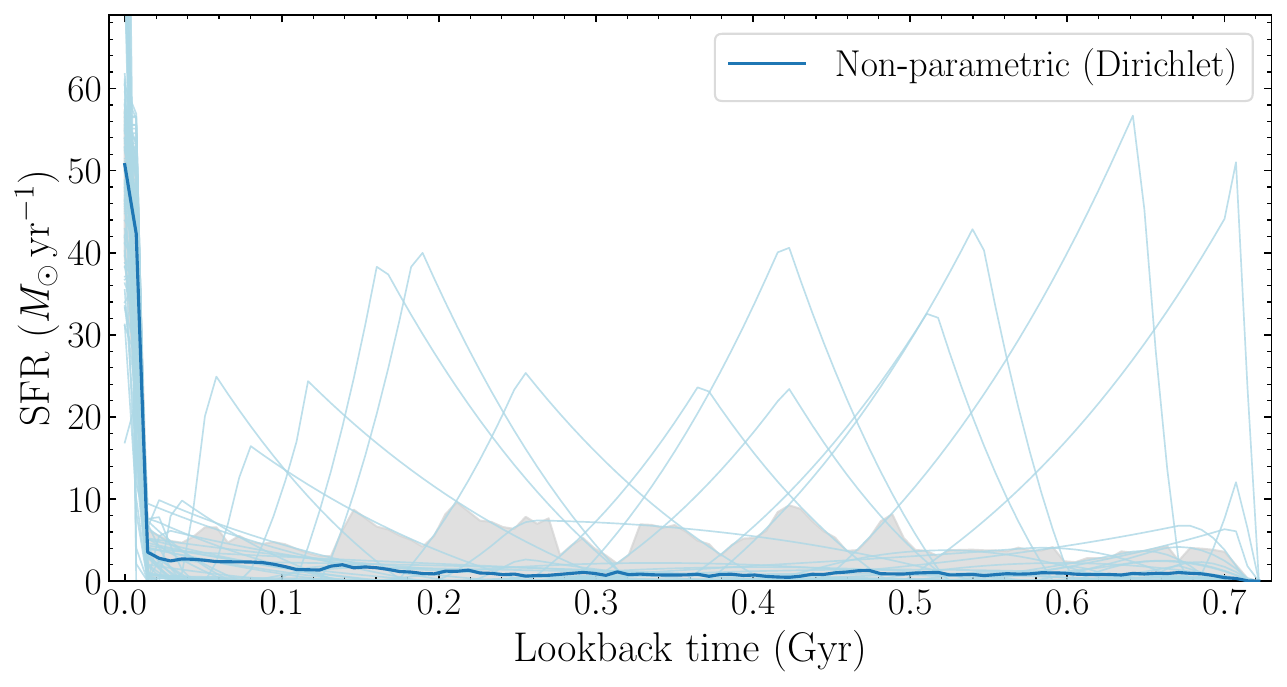}
    \caption{Reconstructed SFH using non-parametric model from \citet{iyer19} and fitted with Dense Basis. The light blue lines are 100 posterior SFHs, while the deep blue line and grey-shaded regions are the 50th and 16th-84th percentile distributions, respectively.}
    \label{iyer_sfh}
\end{figure}

Fig. \ref{sfh} shows the SFHs reconstructed by BAGPIPES fittings. Fig. \ref{iyer_sfh} shows the SFH reconstructed by Dense basis fitting.
\label{sed_appendix}
\section{Details of galaxies plotted in Fig. 8 and Fig. 9}
\label{alma_detected_appendix}
For JWST-selected galaxies in Fig. \ref{age_mass}, the ages are light-weighted ages derived from BEAGLE SED fitting with a constant SFH (CSFH) uniformly \citep{Endsley23faintuv, tang23, chen23, endsley_jades}, reflecting the ages of young populations with recent bursts. However, \citet{endsley_jades} found that adopting a non-parametric SFH with a continuity prior in PROSPECTOR could yield approximately 0.8 dex (for age$_{\rm{CSFH}}<10$ Myr) or 0.3 dex (for age$_{\rm{CSFH}}>100$ Myr) larger stellar mass compared to CSFH-derived mass. This may suggests that the CSFH-derived mass represents a lower limit, or the mass of young stellar populations, while the mass from non-parametric continuity SFH implies an upper limit, including potentially present older stellar populations. Therefore, we plotted mass from CSFH in the left panel and mass from non-parametric continuity SFH in the right panel in Fig. \ref{age_mass}, for comparison. The CEERS samples at $5 < z < 7.5$ lack information of continuity-derived mass \citep{chen23}, thus they are only plotted in the left panel of Fig. \ref{age_mass}.

In Fig. \ref{mass_Z}, the JWST-selected galaxies are compiled from different papers with spectroscopic observations. The JADES samples are from \citet{curti24} whose metallicities are derived by strong-line methods. The CEERS, ERO and GLASS samples are from \citet{nakajima23}, where the metallicities of most galaxies are measured using strong-line methods, while others are determined by direct $T_e$ method. We excluded samples with AGN features.

We summarize the galaxies detected by both ALMA and JWST in Table \ref{alma_detected_table}. For the mass plotted in Fig. \ref{age_mass}, (1) if there are estimates of underlying old stellar mass in the literature, we plot the mass of young populations in the left panel, and the total mass (including potential old populations) in the right panel; (2) if there are consistent results from different SFHs or SED fitting codes, or there are results from only one SED fitting procedure, we use the fiducial values from the corresponding literature in both panels. Similarly, in Fig. \ref{mass_Z}, (1) we plot both young and total mass. The total mass is enclosed in a rectangle, and the two masses are connected by a black dashed line; (2) we plot the fiducial mass values.
\begin{table*}
    \centering
    \caption{Summary of properties of ALMA-detected galaxies discussed in Section \ref{evolutionary pathway}. Unless otherwise specified, metallicity and $M_{\rm{UV}}$ are from the corresponding literature listed below for each target. All values are corrected for gravitational magnification if necessary.}
    \label{alma_detected_table}
    \begin{threeparttable}
        \renewcommand{\arraystretch}{1.3}
        \begin{tabular}{cccccccc}
        \hline
             & $z$ & $\log(M_\ast^{\rm{young}}/\rm{M}_\odot)$ & $\log(M_\ast^{\rm{total}}/\rm{M}_\odot)$ & Age (Myr) & $M_{\rm{UV}}$ & $12 + \log(\rm{O/H})$\tnote{$^\dagger$} \\
        \hline
        SXDF-NB1006-2\tnote{$^a$} & 7.212 & $8.58^{+0.05}_{-0.04}$ & $<9.55$ & $1.8^{+0.1}_{-0.2}$ & -22.1 & $8.07\pm0.12$ \\
        B14-65666\tnote{$^b$} & 7.15 & $9.60^{+0.25}_{-0.29}$ & $<10.0$ & $15^{+17}_{-9}$ & -22.5 & $8.15^{+0.07}_{-0.08}$ \\
        COS-2987\tnote{$^c$} & 6.808 & $8.79\pm0.03$ & $<9.78$ & $8.3^{+2.0}_{-1.2}$ & -21.9 & $7.54\pm0.15$ \\
        MACS0416-Y1\tnote{$^d$} & 8.31 & $8.89^{+0.03}_{-0.01}$ & $9.51\pm0.09$ & $4.76^{+0.28}_{-0.35}$ & -20.8 & $7.76\pm0.03$ (direct $T_e$) \\
        S04590\tnote{$^e$} & 8.496 & $7.15\pm0.15$ & $7.60\pm0.40$ & $ 1.3_{-0.3}^{+0.4}$ & -18.0 & $7.16_{-0.12}^{+0.10}$ (direct $T_e$) \\
        \hline
             & $z$ & \multicolumn{2}{c}{$\log(M_\ast^{\rm{fiducial}}/\rm{M}_\odot)$} & Age (Myr) & $M_{\rm{UV}}$ & $12 + \log(\rm{O/H})$\tnote{$^\dagger$} \\
        \hline
        COS-3018\tnote{$^f$} & 6.854 & \multicolumn{2}{c}{$9.8\pm0.1$} & $18.9^{+5.4}_{-8.7}$ & -22.0 & $8.05\pm0.02$ \\
        Himiko\tnote{$^g$} & 6.595 & \multicolumn{2}{c}{$9.2\pm0.1$} & $8.7^{+2.1}_{-1.6}$ & -22.1 & $7.95\pm0.10$ (direct $T_e$) \\
        MACS1149-JD1\tnote{$^h$} & 9.114 & \multicolumn{2}{c}{$8.21^{+0.06}_{-0.07}$} & $\sim10$ & -19.3 & $7.90^{+0.04}_{-0.05}$ (direct $T_e$) \\
        A2744-YD4\tnote{$^i$} & 7.88 & \multicolumn{2}{c}{$9.17_{-0.17}^{+0.27}$}  & $36.3_{-25.6}^{+142}$ & -19.7 & $8.19^{+0.09}_{-0.13}$ \\
        A2744-YD7\tnote{$^i$} & 7.88 & \multicolumn{2}{c}{$9.39^{+0.08}_{-0.15}$} & $195^{+129}_{-87.8}$ & -20.0 & $7.53^{+0.37}_{-0.49}$ (SED fitting) \\
        JADES-GS-z14-0\tnote{$^j$} & 14.1793 & \multicolumn{2}{c}{$8.7^{+0.5}_{-0.2}$} & $15^{+28}_{-12}$ & -20.8 & $7.69\pm0.30$ (Cloudy modeling) \\
        GHZ2\tnote{$^k$} & 12.33 & \multicolumn{2}{c}{$8.92_{-0.28}^{+0.13}$} & $28^{+10}_{-14}$ & -20.5 & $7.40^{+0.52}_{-0.37}$ \\
        CR7\tnote{$^l$} & 6.60 & 
        \multicolumn{2}{c}{$9.29^{+0.17}_{-0.16}$} & ... & -22.2 & $8.0\pm0.5$ (direct $T_e$) \\
        A2744-YD1\tnote{$^m$} & 7.88 & \multicolumn{2}{c}{$8.74\pm0.19$} & ... & -19.0 & $8.05_{-0.06}^{+0.04}$ \\
        BDF-3299\tnote{$^n$} & 7.11 & \multicolumn{2}{c}{$8.45_{-0.12}^{+0.31}$} & ... & -20.4 & $7.72_{-0.13}^{+0.07}$ \\
        COSMOS24108\tnote{$^o$} & 6.36 & \multicolumn{2}{c}{$9.53_{-0.09}^{+0.10}$} & ... & -21.7 & $8.21_{-0.05}^{+0.03}$\\
        REBELS\tnote{$^p$} & $\sim6.5-8$ & \multicolumn{2}{c}{$\sim9.2-9.8$} & ... & $\sim-21.2--22.4$ & $\sim7.8-8.7$ \\
        \hline
        \end{tabular}
        \begin{tablenotes}
            \item Note:
            \item[$^\dagger$] The parentheses behind the metallicities indicate what method they are derived from. Metallicities without parentheses behind are measured from strong-line methods.
            \item References: 
            $a$: The target of this work. $M_\ast^{\rm{young}}$ and age are from lognormal SFH. $M_\ast^{\rm{total}}$ is the sum of young mass and maximum potential old mass.
            $b$: $M_\ast^{\rm{young}}$ and age are from CSFH without a parameter $M_{\rm{old}}$. $M_\ast^{\rm{total}}$ is the sum of young mass and maximum potential old mass derived from CSFH with a parameter $M_{\rm{old}}$ \citep{sugahara25}. Metallicity is calculated from \citet{Jones24} assuming $12+\log(\rm{O/H})_\odot = 8.69$ \citep{asplund2001}.
            $c$: $M_\ast^{\rm{young}}$ and age are determined through CSFH. The upper uncertainty of $M_\ast^{\rm{young}}$ is assumed to be the same as the lower uncertainty. $M_\ast^{\rm{total}}$ is the upper limit of mass including hidden old populations \citep{mawatari25}.
            $d$: $M_\ast^{\rm{young}}$ and age are measured from $\tau$ SFH \citep{ma24}. $M_\ast^{\rm{total}}$ is the sum of four components \citep{Harshan24}.
            $e$: $M_\ast^{\rm{young}}$ and age are from CSFH \citep{Heintz_z8}. We adopted the conservative mass assumption in \citet{Heintz_z8} for $M_\ast^{\rm{total}}$. $M_{\rm{UV}}$ is from \citet{nakajima23}.
            $f$: Mass, age and $M_{\rm{UV}}$ are from \citet{harikane25}. Metallicity is the average of three components when considering [N\,{\sc ii}]$\lambda6585$/H$\alpha$ ratio \citep{scholtz24}.
            $g$: Mass, age and $M_{\rm{UV}}$ are from \citet{harikane25}. Metallicity is the average of three components \citep{kiyota25}.
            $h$: All information is from \citet{stiavelli23}.
            We adopt the age of the recent major burst.
            $i$: Mass and age are from \citet{morishita23}. The metallicities of YD4 and YD7 are from \citet{venturi24} and \citet{morishita23}, respectively. 
            $j$: Mass and age are from \citet{Helton25}. Metallicity is from \citet{schouws25}, in which the fiducial value is $Z\sim0.05-0.2\,\rm{Z}_\odot$. We converted this value to $12+\log(\rm{O/H})=7.69\pm0.30$ assuming $12+\log(\rm{O/H})_\odot = 8.69$ \citep{asplund2001} and its $1\,\sigma$ uncertainty has the same values.
            $k$: All data are from \citet{zavalaMIRI}.
            $l$: Mass and metallicity are from \citet{Marconcini25}. $M_{\rm{UV}}$ is from \citet{harikane25}.
            $m$: Mass and $M_{\rm{UV}}$ is from \citet{Witten25}. Metallicity is the average of two components in \citet{venturi24}.
            $n$: Mass is the sum and metallicity is the average of three components in \citet{venturi24}. $M_{\rm{UV}}$ is from \citet{Binggeli21}.
            $o$: Mass is the sum and metallicity is the average of three components in \citet{venturi24}. $M_{\rm{UV}}$ is from \citet{harikane25}.
            $p$: Mass and metallicity are from \citet{Rowland25}. $M_{\rm{UV}}$ is adopted from \citet{fisher25}.
        \end{tablenotes}
    \end{threeparttable}
\end{table*}

% Don't change these lines
\bsp	% typesetting comment
\label{lastpage}
\end{document}